\documentclass[usenatbib]{mn2e}

\usepackage{epsfig}
\usepackage{colordvi}
\usepackage{html}

\newcommand{\HI}{\mbox{\rm H\,\footnotesize{I}}}

\newcommand{\nid}{\noindent}
\newcommand{\rtp}[1]{\ensuremath{^{#1}}}
\newcommand{\abs}[1]{\ensuremath{\mid #1 \mid}}
\newcommand{\apx}{\ensuremath{\sim}}
\newcommand{\pmt}{\ensuremath{\pm}}

\newcommand{\Vr }{\ensuremath{V_{r  }}}

\newcommand{\Vx }{\ensuremath{V_{x  }}}
\newcommand{\Vy }{\ensuremath{V_{y  }}}
\newcommand{\Vyp}{\ensuremath{V_{y' }}}

\newcommand{\mumC}{\ensuremath{\mu_{m}^{C}}}
\newcommand{\muMC}{\ensuremath{\mu_{M}^{C}}}
\newcommand{\mumCsq}{\ensuremath{(\mu_{m}^{C})^2}}
\newcommand{\muMCsq}{\ensuremath{(\mu_{M}^{C})^2}}

\newcommand{\Myp}{\ensuremath{\mu_{y'}}}
\newcommand{\Mx}{\ensuremath{\mu_{x}}}

\newcommand{\muas}{\ensuremath{\mu\mbox{as}}}

\newcommand{\muasyr}{\ensuremath{\mu\mbox{as\,yr}^{-1}}}

\newcommand{\masyr}{\ensuremath{\mbox{mas} \, \mbox{yr}^{-1}}}
\newcommand{\kms}{\ensuremath{\mbox{km\,s}^{-1}}}
\newcommand{\kmsyr}{\ensuremath{\mbox{km\,s}^{-1} \,\mbox{yr}^{-1}}}

\newcommand{\kmsMpc}{\ensuremath{\mbox{km\,s}^{-1}\,\mbox{Mpc}^{-1}}}

\newcommand{\ph} [1]{\phantom{#1}}

\newcommand{\GAIA}     {{\em GAIA}}
\newcommand{\HIPPARCOS}{{\em HIPPARCOS}}
\newcommand{\HST}      {{\em HST}}
\newcommand{\PLANCK}   {{\em PLANCK}}
\newcommand{\SIM}      {{\em SIM}}
\newcommand{\OBSS}     {{\em OBSS}}
\newcommand{\SKA}      {{\em SKA}}
\newcommand{\WMAP}     {{\em WMAP}}

\newcommand{\aap}  {A\&A}
\newcommand{\aaps} {A\&AS}
\newcommand{\aj}   {AJ}
\newcommand{\apj}  {ApJ}
\newcommand{\apss} {Ap\&SS}
\newcommand{\araa} {ARA\&A}
\newcommand{\mnras}{MNRAS}
\newcommand{\pasp} {PASP}
\newcommand{\prd}  {PhRvD}

\newcommand{\eg}{{\it e.g.},\ }
\newcommand{\etal}{{\it et al.}}

\newcommand{\snROv}[1]{\renewcommand{\baselinestretch}{#1}\begin{normalsize}}
\newcommand{\enROv}{\end{normalsize}\renewcommand{\baselinestretch}{1.0}}

\newcommand{\ssROv}[1]{\renewcommand{\baselinestretch}{#1}\begin{small}}
\newcommand{\esROv}{\end{small}\renewcommand{\baselinestretch}{1.0}}

\begin{document}

\title[Accurate Extra-Galactic Distances and Dark Energy]
      {Accurate Extra-Galactic Distances and Dark Energy:\\
       Anchoring the Distance Scale with Rotational Parallaxes}

\author[Rob P. Olling]
       {Rob P. Olling$^1$\thanks{email: olling@astro.umd.edu}\\
        $^1$Astronomy Department, University of Maryland, 
         College Park, MD 20742-2421
	}

\maketitle
\begin{abstract}
%
%
We investigate how the current and future uncertainty on the Hubble
constant affects the uncertainty in the Equation of State of Dark
Energy ($w$) and the total density of the Universe
($\Omega_{tot}$). We start with the approximate linear relations
between the cosmological parameters as presented by \citep{WMAP07},
and use the standard error-propagation relations to estimate the
effects of improving the Cosmic Microwave Background (CMB) parameters
as well as the Hubble constant ($H_0$) on our knowledge of the
Equation of State (EOS) of Dark Energy. Because we do not assume a
flat universe, we also estimate the attainable accuracy on
$\Omega_{tot}$ and the spatial curvature of the Universe.

In one limiting case, we assume that the constraints provided by
additional data (galaxy clustering, weak lensing and so forth) do not
improve significantly, while the error on the Hubble constant is
decreased by a factor up to ten.  The other limiting case of
significantly improved additional data with current $H_0$ errors has
been investigated by the Dark Energy Task Force [DETF;
\citet{DETF2006}]. For the former scenario, we find that future
improvements of the determination of the CMB hardly changes the
accuracy with which the EOS and $\Omega_{tot}$ are know, {\em unless}
the Hubble constant can be measured with an accuracy of several
percent. The conclusion of the DETF is that the Hubble constant hardly
matters if the additional data is sufficiently accurate.  We find that
a combination of moderate $H_0$ improvements with moderately improved
``other'' data might significantly constrain the evolution of dark
energy, but at a reduced cost.

We review in some detail several methods that might yield
extra-galactic distances with errors of order of several percent,
where we focus on the current and future strengths and weaknesses of
the methods. Specifically we review the following: the Velocity Field
method, two Maser methods, four Light Echo techniques, two Binary Star
methods, and the ``Rotational Parallax'' (RP) technique. Because these
methods substantially rely on geometry rather than astrophysics or
cosmology, their results are quite robust.

In particular we focus on the advantages of the Rotational Parallax
technique which can provide accurate (1\%), single-step, and bias-free
distances to Local Group galaxies. These distances can be used to
improve the zero-point for other distance indicators which in turn
would then be able to determine the Hubble constant to greater
accuracy than they currently do.  Achieving an accuracy of a few
percent in the zero-point distances to M\,31, M\,33 and the LMC by the
RP method requires radial velocities at the 10 \kms\ level and proper
motions attainable by future astrometric missions such as \SIM, \GAIA\
and \OBSS, or by future radio observatories such as the \SKA.

\end{abstract}

  \section{Introduction}
\label{sec:Introduction}
A Key Project of the Hubble Space Telescope (\HST) resulted in a
fairly accurate determination of the Hubble constant: $H_0$ = 74 $\pm$
2 (random) $\pm$ 7 (systematic) \kmsMpc\ \citep{H0HST}. There are a
number of sources of systematic error, but the largest contribution to
the systematic error is due to the uncertainty in the adopted distance
to the Large Magellanic Cloud (\pmt\ 6.5\%). Notwithstanding the
systematic uncertainty, relative distances measurements to galaxies
can now be made with a precision (not accuracy\footnote{In this paper
we make a distinction between precision and accuracy, in the sense
that the former refers to small internal errors, while the latter
indicates small external errors. }) of 5-10\%.

The drive towards accurate cosmological parameters is in part
motivated by the connection with fundamental physics theories that
might be able to predict observables such as the value of the critical
energy-density of the Universe $\rho_{crit} \equiv 3 H_0^2 / (8\pi G)$
[e.g., \citet{C2001,DETF2006} and many others].  If one assumes a flat
$\Lambda$CDM model, then the fluctuations of the cosmic microwave
background (CMB) as observed by \WMAP\ imply a similar value and
uncertainty of $H_0$ [=73 $\pm$ 3, \citet{WMAP03,WMAP07}; hereafter
referred to as WMAP07] as the \HST\ results by \citet{H0HST}. If the
flatness assumption is abandoned, \WMAP\ hardly constrains $H_0$
because the \WMAP\ data constrains the actual matter density: that is
to say, the product of the normalized matter density ($\Omega_m$) and
$h^2$ (WMAP07, their figure 20), where $h \equiv 100 \, \kmsMpc /
H_0$, and $\Omega_m h^2 \equiv \omega_m$ (see
\S\ref{sec:Implications_for_Cosmology}). Obviously, one way to
determine $\Omega_m$ is to accurately determine the Hubble constant.

Another way to determine $\Omega_m$, the Hubble constant and other
cosmological parameters is to employ ``other'' data sets that exhibit
different dependencies on $h$. Such a method is preferred while the
error on $h$ is of order a factor two. The constraints from other
data sets used by WMAP07 are: large-scale structure observations
galaxy redshift surveys), distant type-Ia Supernovas, Big-Bang
nucleosynthesis, Sunyaev-Zel'dovich (SZ) fluctuations, Lyman-$\alpha$
forest, and gravitational lensing. However, the physical processes
underlying these constraints can be more complex than those of the CMB
and/or ``geometric'' determinations of the Hubble constant. Reliance
on these additional data can potentially lead to serious biases in the
derived cosmological parameters [e.g., \citet{Sea2003, PLANCK}].
Because it is beginning to become possible to achieve highly accurate
``geometric'' determinations of $H_0$, it will soon be more
profitable to use $H_0$ as an independent constraint rather than as a
to-be-fitted parameter.

Combination of WMAP and the other data sets yields strong evidence
that the Universe is indeed close to being flat with a total
normalized density of $\Omega_{tot} = 0.996 \pm 1.6\%$ (see section
\ref{sec:Implications_for_Cosmology} below). Combining the \WMAP\ data
with the Hubble constant yields the equation of state ($w$) of Dark
Energy (DE): $w = -0.95 \pm 12\%$ [eqn.~(\ref{eqn:w_now_rel})], where
the equation of state (EOS) is the ratio of pressure ($p$) and density
($\rho$). This results follows from the assumption of a flat
universe. If flatness is {\em not} assumed, our approximation yields
$w -0.98 \pm 7.2\%$ [eqn.~(\ref{eqn:Delta_w_omega_h})], which is
close to the rigorously derived value ($w = -1.062_{-0.079}^{+0.128}$)
of \citet{WMAP07}. In this paper we explore how the uncertainties on
$\Omega_{tot}$ and $w$ depend on the uncertainty of the Hubble
constant. We develop our models for the case that $w$ does not evolve,
and that the accuracies of the other data sets do not improve. We then
re-scale these models to take $w$-evolution and data-quality
improvements into account.

We find that accurate knowledge of these parameters requires better
CMB data (as provided by \PLANCK), but also significantly smaller
errors on $H_0$ (\S\ref{sec:Implications_for_Cosmology}). In contrast,
the Dark Energy Task Force (DETF) finds that the errors on $H_0$
hardly matter as long as the other data sets improve significantly
in quality [\citet{DETF2006}, and
\S\S\ref{sec:Comparison_with_the_results_from_the_DETF}]. We suggest
that DETF-Stage~IV Dark-Energy knowledge might be achieved with less
focus on large-scale surveys of the distant Universe by incorporating
a highly accurate measure of $H_0$ obtained at earlier stages such as
those provided by the ``Water Maser Cosmology Project''
\citet{WMCP2006} and/or independently by a rotational-parallax program
\S\ref{sec:Rotational_Parallax_Distances}. If that is indeed the case,
substantial cost and time savings might be achieved. Of course, a
locally determined $H_0$ value also serves as a completely
independent check on ``traditional'' CMB-based $H_0$ determinations
such as WMAP07. Also, future astrometric missions will provide
accurate constraints on the minimum age of the Universe as obtained
from binary stars (\S\ref{sec:Discussion_and_Conclusions}).

This paper is organized as follows. In section
\ref{sec:The_Extra_Galactic_Distance_Scale} we try to summarize the
current state of the zero-point of the extra galactic distance scale,
with an emphasis on methods that may be able to yield extra-galactic
distances at the percent level
(\S\S\ref{sec:Direct_Distance_Measures}). In the appendix, we
work out that distance estimates based on water-masers may be biased
if elliptical orbits are present in AGN disks. In section
\ref{sec:Implications_for_Cosmology} we discuss our modeling procedure
to estimate the errors on the equation of state of Dark Energy as a
function of the errors on the Hubble constant. Section
\ref{sec:Rotational_Parallax_Distances} describes the Rotational
Parallax method in detail, where we also differentiate between the
abilities between the on-going and planned astrometric missions:
\GAIA, \SIM\ and \OBSS. We discuss and summarize our results in
section \ref{sec:Discussion_and_Conclusions}. Here we also describe
other (future) astrometric constraints on cosmology.
\vspace*{-2.2em}
  \section{The Extra-Galactic Distance Scale}
\label{sec:The_Extra_Galactic_Distance_Scale}
During the past decades significant progress has been made on the
calibration of the extra-galactic distance scale, and the
determination of the Hubble constant with methods such as type Ia
Supernovas, the Tully-Fisher relation, surface brightness
fluctuations, the Tip of the Red-Giant Branch (TRGB) and the
Fundamental Plane.  These distance-determination techniques are now
relatively free of systematic uncertainties. However, the primary
calibration of these methods comes from the period-luminosity relation
for Cepheid variables.

However, the true value of zero-point of the Cepheid distance scale is
still debated, in particular the difference between the zero-points
based on Cepheids in the Large Magellanic Cloud (LMC) and those in the
Milky~Way [\eg\ \citet{FC1997, MF1998, GO2000, B2002, R1999}]. Of the
Cepheids with \HIPPARCOS\ measurements \citep{ESA97}, all but 2 have
signal-to-noise (S/N) values less than five \citep{FC1997}, so that
systematic biases (\eg Lutz-Kelker correction) are important.  Thus,
the calibration of the Cepheid period--luminosity relation relies on
Galactic Cepheids in open clusters, and is therefore tied to
main-sequence fitting for clusters calibrated to the Hyades, and the
till-recently-problematic Pleiades \citep{Sea2005}.  However, steady
progress is being made in the calibration of Galactic Cepheids
distances via \HST\ trigonometric parallaxes
[\citet{Bea_HST_Cep2002_2007}] and interferometric calibration of the
Baade-Wesselink method \citep{Lea2002, Nea2002, Kea2004a, Kea2004b}.

Because of uncertainties in the Galactic calibration, most
extra-galactic distance scale studies have been calibrated relative to
the nearby LMC. However, the metallicity of the LMC is substantially
below that of those distant galaxies (and the Milky~Way [MW]) that are
used to calibrate the Supernova Ia distance scale onto the Cepheid
distance scale [e.g., \citet{Sand2006}], while there are also strong
indications that the Period-Luminosity (-Color) relation is non-linear
[e.g., \citet{NK2006,Groen2004} and references therein] and dependent
on metallicity [e.g., \citet{G1994, Sea1997, Sea2004, Groen2004,
Sand2006}].  In fact, the range in recently-published values of $H_0$
(58 -- 75 \kmsMpc) may be entirely attributable to differences in the
applied metallicity corrections \citep{Sand2006}.

Several independent methods for measuring distances are commonly
applied to the LMC.  These include Cepheids, the Red Clump, eclipsing
binaries, SN1987A, the magnitude of the TRGB, RR~Lyraes, and Miras.
Unfortunately, the full range of most of the distance modulus
($\mu_{LMC}$) to the LMC extends from 18.2 to 18.7 mag \citep{R1999}.
The wide range of moduli relative to the quoted internal errors
indicates that systematic errors still dominate the determinations of
the LMC distance. \citet{A2004} finds that recent applications of a
variety of methods seem to settle on $\mu_{LMC}=18.5$ \pmt\ 0.1 mag
(or $H_0$ = 71 \pmt\ 10 \kmsMpc), which corresponds nicely with the
\WMAP\ value.  However, it is possible that such convergence is partly
related to (over-) confidence in the \WMAP\ results.  After all, some
inconsistencies remain. For example, the K-band Red-Clump distance to
the Galactic centre equals 8.24 \pmt\ 0.4 kpc \citep{A2000} while
geometric methods yield values one to two sigma smaller, or about 7.3
kpc: 1) the ``expanding water maser'' distance to Sgr B2 equals 7.2
\pmt\ 0.7 kpc \citep{R1993}, and 2) the bias-free ``orbital parallax
method'' \citep{Aea1992} of stars orbiting Sgr A\rtp{*} yields 7.4
\pmt\ 0.2 kpc (Ghez, 2006, private communications).

The TRGB method \citep{DCA1990, LFM1993} is based on a break in the
luminosity function of the ``giant branch,'' which is due to to the
maximum attainable luminosity before the helium flash [e.g.,
\citet{IR1983,SC1997}]. The TRGB method is independent of the adopted
distance to the LMC because it uses Galactic Globular Clusters (and
hence RR~Lyrae\footnote{\citet{BFP2001} employ a distance to $\omega$
Cen based on a double-line spectroscopic binary.}) as calibrators
instead. For metal-poor systems, this method has a metallicity
dependence of approximately $M^{TRGB}_{I} \approx -3.66 + 0.48
[Fe/H] + 0.14 [Fe/H]^2$ \citep{BFP2001} which is fairly small for
$[Fe/H] \la -0.6$, but increases rapidly towards higher
metallicities. If the metallicity can be determined
independently\footnote{In practice, the metallicity is often estimated
from the photometry.} the random errors of the TRGB method correspond
to a distance error of several percent, while the systematic error of
approximately 10\% is mostly due to the uncertainties in the Globular
Cluster distance scale [see, for example, \citet{Mea2005, RBKGP2006}
for recent discussions].  Thus, the TRGB will provide accurate
extra-galactic distances as long as a sufficient number of RGB stars
[$\ga 50$, e.g., \citet{MF1995, MMRTDSS2006}] can be resolved, and the
metallicity can be estimated accurately. However, given that
galaxy-halos are often the target for TRGB programs and that these
halos often have large spread in metallicity \citep{DHP2001,
MRFBS2005, M2006}, metallicity effects will be important for very
accurate TRGB results. Nevertheless, a large majority of accurate
galaxy distances with $D \la 12$ Mpc currently derives from the TRGB
method \citep{Kea2006}. In fact, the TRGB method may be the most
robust photometric method available for distance determination in the
nearby Universe.

\vspace*{1em}

Although it would be preferable if the first steps of the distance
ladder could be avoided altogether via trigonometric parallaxes, such
measurements are beyond the capability of future astrometric missions
such as NASA's \SIM\ PlanetQuest\footnote{
http://planetquest.jpl.nasa.gov/documents/\\WhitePaper05ver18\_final.pdf}
[\eg \citep{SIM05}], ESA's \GAIA\footnote{
http://astro.estec.esa.nl/SA-general/Projects/GAIA/gaia.html}
\citep{P2002} and the proposed \OBSS\ mission \citep{OBSS}. The
accuracy of \SIM's grid is expected to be about 3.5 \muas, which
corresponds to a distance error of about 19\% for the LMC, while for
\GAIA\ and \OBSS\ the accuracy is coarser by a factor of two. Note that
\SIM's status is currently unclear\footnote{While it is ``placed on
hold'' there is 20 -- 30 M\$ (US) per year in NASA's budget for the
\SIM\ mission.}. The Square Kilometer Array
[\SKA\footnote{http://www.skatelescope.org/; The \SKA\ is planned to
be fully operational around 2020.}; \citet{SKA_2004}] will have
astrometric capabilities \citep{FR2004,F2005} comparable to
\GAIA\ and \OBSS\ for those sources that have strong radio
emission, and will thus not be able to measure extra-galactic
parallaxes directly.

Often, many members of ``standard candle'' group are located within
several kpc from the Sun. Most of these standard-candle methods
(supernovae being an obvious exception) will be revitalized with the
advent of the next generation space astrometry missions. Also, it is
possible to obtain rotational parallaxes at the required accuracy
required for external galaxies distance determination [\citet{PS1997}
and \citet{OP2000, OP2007}, hereafter referred to as OP2000]. This is
discussed in \S~\ref{sec:Rotational_Parallax_Distances} below.
\subsubsection{``Nearby'' Extra-Galactic Calibrators}
      \label{sec:Nearby_Calibrators}

A future determination of an accurate distance to a single galaxy does
not amount to an accurate determination of the Hubble constant. In
most cases, the Hubble constant would need to be determined via
secondary calibrators in galaxies at larger distances because typical
deviations from the linear Hubble flow due to interactions with
companions, groups, (super)clusters and voids do occur. For nearby
galaxies, the size of these effects virtually guarantee that the
measured recession velocity of the calibrator is atypical for the
distance of the galaxy. If a typical deviation from the Hubble flow
is, say, 200 \kms, then a 1\% accuracy of the Hubble constant would
require a calibrator galaxy at redshift 200/0.01 = 20,000 \kms, or
about 280 Mpc to eliminate the uncertainty induced by non-linear
Hubble flow.

Currently, the only galaxy with a distance error of \apx 5\% is
NGC~4258. At a distance of 7.2 Mpc \citep{Hea1999, Hea2005} has just
begun to serve as a calibrator for secondary distance indicators: the
TRGB method \citep{Mea2005} and the Cepheid
Period-Luminosity-Abundance relation \citep{Nea2001, Mea2006}.
However, a closer calibrator with at least comparable accuracy is
preferable because secondary effects due to metallicity and so forth
can be studied and corrected-for on a star-by-star basis.  The LMC,
M\,31 and M\,33 are the obvious targets, while the rotational parallax
method is the ideal method to obtain unbiased distances [see
\S~\ref{sec:Rotational_Parallax_Distances}].

Alternatively, one can use a larger number of less accurate nearby
calibrator galaxies. If Cepheids are used as secondary distance
indicators, the resulting distance error might not decrease much below
5\% due to metallicity effects as well as uncertainties in the Helium
abundance and other peculiarities of individual Cepheids (I thank the
referee for pointing this out). On the other hand, it is quite
possible that these individual peculiarities can be calibrated to
better accuracy with highly accurate parallax and photometric \GAIA\
data for many hundreds of Galactic Cepheids, TRGB stars, RR~Lyrae and
so forth. 

\vspace*{-1.5em}
\subsection{``Direct'' Distance Measures}
 \label{sec:Direct_Distance_Measures}
Fortunately, other more direct methods that can yield reliable
distances have been proposed (and applied) in the literature. These
methods can be applied in either the optical and/or the radio
regime. Below we review the most promising of the candidates that we
have been able to identify in the literature. Some science projects
require extra-galactic distances with an accuracy of 5\% or 10\% that
will be readily attainable for a large number of galaxies may be
sufficient, while other projects require the very best possible
accuracy. The TRGB method might be applied for projects of the former
type as soon as its zero-point has been accurately established. The
review below is focussed on the latter type: methods that can deliver
distance accuracies at the percent level.
\vspace*{-1.5em}
\subsubsection{Velocity Field Methods}
    \label{sec:Velocity_Field_Methods}
The ``Rotational Parallax'' (RP) method was conceived by
\citet{PS1997} in the \SIM\ context as an extra-galactic variant of
the ``orbital parallax'' method. In the latter method, radial
velocities and proper motions of a resolved binary system are combined
to yield all the orbital parameters, as well as the distance [\eg,
\citet{Aea1992,Dea2005}]. In the simplest extra-galactic application,
one assumes that: 1) the orbits are close to circular, 2) with
deviations only due to (small) random motions, 3) the inclination can
be obtained from the analysis of the radial velocity field, and 4) the
dynamical centre is known from the velocity field.  With these
simplifying assumptions and the use of just two stars per galaxy, the
systemic motion of the galaxy can be subtracted and the distance
determined from the orbital model and the observed radial and proper
motions. A more detailed treatment can be found in
\citep{OP2000, OP2007} and \S~\ref{sec:Rotational_Parallax_Distances}.
The trigonometric-, orbital- and rotational parallax methods rely only
on geometric relations, Newton's laws of motion (and the ability to
measure accurate positions, proper motions and radial
velocities). These methods are entirely independent of photometric
transformations, metallicities, effective-temperature scales,
pulsation theory and other uncertain properties of stars such as their
interiors and their atmospheres \citep{rK2002}. 

The ``Velocity Gradient'' method is in part geometric, and in part
photometric and has been applied to open clusters and modified to work
for the LMC [see \citet{AG2000}, and references therein]. In this
method, the transverse velocity of the centre-of-mass of the
cluster/galaxy induces an angular difference between the kinematic and
photometric major axes (about 20\rtp{o} for the LMC). In practice this
method would require radial velocity and proper motion measurements of
about 10,000 stars with accuracies better than about 10 \kms\ and 150
\muasyr, respectively, which are easily attainable by \GAIA. The most
difficult part of this method is the determination of the photometric
major-axis: this requires 0.1 mmag photometric accuracies over roughly
10 degrees as well as accurate reddening estimates
\citep{AG2000}. Although it may not be possible to obtain such data
from the ground, \GAIA\ is well-equipped to deliver all the required
data for this method. This method might also be applied to \GAIA\ data
of M~33.  The high inclination of M~31 will likely preclude successful
application of the velocity gradient method. Note that in practice the
Rotational Parallax Method (or a simplified version thereof) may be
needed to determine the systemic proper motion of the LMC [see
\S~\ref{sec:Rotational_Parallax_Distances} and \citet{OP2000}].
\vspace*{-1.5em}
\subsubsection{Maser Methods}
    \label{sec:Maser_Methods}
Water masers associated with star forming regions scattered across the
disk of a spiral galaxy can be used to estimate the distance to that
galaxy. With two masers sources located on ``opposite'' sides of the
centre, the known rotation curve and inclination of the galaxy can be
used to derive the distance. \citet{Bea_H2O_2005} identified 37 maser
features in two masing complexes in M~33 for which they could
determine proper motions. From the per-complex average motion in
combination with the velocity-field model, \citet{Bea_H2O_2005}
determined the distance to M~33 with an error of 23\%.  Note that this
method is a version of the rotational parallax method, but one that
relies on external information [see
\S~\ref{sec:Rotational_Parallax_Distances} and \citet{OP2000}]. 
All the systematic effects that could affect the RP method are also
important for the water-maser work, especially when only a limited
number of sources are available per galaxy [M~31 is expected to have
17 H$_2$O masers at \SKA\ sensitivity \citep{Bea2006_H2O}].  When the
\SKA\ becomes operational, many more fainter maser sources will be
detected in galaxies out to 100 Mpc \citep{F2005}, while a distance
accuracy of 1\% can be achieved out to 30 Mpc \citep{FR2004}. At this
distance, the perturbations in the Hubble flow would allow for a
determination of about 10\% per galaxy: the more distant sources would
allow for the 1\% goal on $H_0$.

Rather than averaging the proper motions per masing region, the H$_2$O
maser data can be utilized in a manner similar to their Galactic
equivalents to derive distances. That is to say: a dynamical model of
the star-forming region is used to predict the relation between the
observable radial velocities and proper motions. Such models can
either be based on ordered motions (such as rotation or expansion) or
on isotropic random motions. See \citet{Aea2004} for a full
description, references and an application to the IC~133 star-forming
complex in M~33. For the case of the 14 year observing campaign of
IC~133, \citet{Aea2004} adopt a random+systematic error of 22\%, which
is dominated by: 1) the unknown centre of expansion (\pmt 16\%), 2)
random error (\pmt 12\%), and 3) an assumed expansion velocity (and
error) of the group of maser spots (as determined by comparison with
Galactic H$_2$O masers; \pmt 9\%) and an assumed systemic velocity
(and error) based on co-incident CO emission (\pmt 4\%). Although the
expended effort is already enormous, it seems likely that the results
could improve by at least a factor of two as a result of continued
long-term monitoring (especially with the \SKA), which would reduce
the RMS errors and would eliminate the secondary possibility for the
expansion centre.

NGC\,4258 is the current record holder for the most accurate
extra-galactic distance, with a total error of about 5\% as derived
from the geometry and dynamics of its nuclear water maser sources
\citep{Hea1999,Hea2005}. The H$_2$O masers are thought to arise in an
almost edge-on circum-nuclear disk in Keplerian rotation. In this
model, the masers occur close to the locations where the line-of-sight
velocity gradients are small: that is to say, close to the major and
minor axes of the velocity field (the major axis is also called the
``midline''). Maser spots close to the major and minor axes are called
the ``high-velocity'' and ``systemic'' masers, respectively.  With the
assumption of circular orbits, the observed positions and velocities
can be converted to a model where the inclination ($i$), the position
angle ($\phi$) and the rotation speed ($V_c$) depend on radius ($R$)
only [see \citet{HMGT2005} and references therein]. Given this model,
the distance can be derived in two independent ways from the systemic
masers.  First, the circular speeds of the systemic masers are
directly observable as tangential proper motion with a value of about
30 \muasyr. Because phase-referencing techniques are used to obtain
the VLBA positions, absolute position information is lost
\citep{HMGT2005}, so that the systemic proper motion of the galaxy is
effectively subtracted from the observations. Thus, the above
relations amount to a specialized version of the rotational parallax
technique (see \S\ref{sec:Rotational_Parallax_Distances}). The
conversion from observed position to radial coordinate $R$ introduces
the largest systematic uncertainty in distance because $V_c$, $i$ and
$\phi$ change fairly rapidly with radius. The second technique employs
the change in radial velocity ($V_r$) that occurs due to the orbital
motion of the systemic masers.  In almost edge-on geometry and with
$V_c \sim 1100$ \kms\ and $P$ about 750 yr, the accelerations are
about 9 \kmsyr. These two geometric techniques yield consistent
results.
\vspace*{-1.5em}
\subsubsection{Nuclear Water Maser: Eccentric Orbits}
    \label{sec:Nuclear_Water_Masers:_Eccentric_Orbits}
A systematic uncertainty that has hitherto not been investigated is
that of elliptical orbits in the masing region.  It is well
established that the majority of the high-luminosity cousins of
NGC\,4258 often exhibit significant degrees of non-axisymmetry in
their disks \citep{Sea_Ell_AGN_2003}. Furthermore, models employing
disks with substantial ellipticity were very successful in describing
the observed optical double-peaked line profiles of AGN
\citep{ELHS1995}, who infer orbital eccentricities of order
three-tenths. Such large eccentricities could result if the AGN
contains in fact a binary black hole, or the tidal disruption of a
normal star for the case of a single black hole. In such models, it
seems likely that the water-masing region would also be affected
because the radio emission is generated just a few times further out
than the optical emission lines. However, other explanations exist for
the optical emission lines, and it is not clear that elliptical orbits
do in fact occur in accretion disks \citep{SB_Ell_AGN_2003}.  In fact,
models with such large and persistent eccentricities have been ruled
out on the basis of long-term spectroscopic monitoring of several AGN
\citep{GHE2007}.  Nevertheless, it seems that the possibility of
elliptical orbits should be taken into account in the systematic-error
budget of nuclear water masers, because eccentricities thirty times
smaller than those ruled out by \citet{GHE2007} would still bias the
inferred distances (see below). Observations of other maser lines (if
present) which probe different regions of the accretion disk would
also minimize the uncertainty associated with large-scale non-circular
motions.

Here we analyze a simple version of the general case of elliptical
orbits where the major axis of the ellipse is aligned with the line
of sight and show that this geometry yield observables that are
indistinguishable from the case of circular orbits, albeit for a
different distance, inclination and mass. For illustrative purposes we
also assume that the disk is not warped. This case is worked out in
some detail in Appendix~\ref{apx:elliptical_orbits}, where we find the
following relation the distance [eqn.~(\ref{eqn:D_ell})]:
$
D_c / D_t = \sqrt{ (1 \mp e)^3 / (1 \pm e) } \, \approx \, 
          1 \mp 2\,e
$,
where $D_c$ is the distance that would be inferred under the
assumption of circular orbits, $D_t$ the true distance as derived for
non-circular orbits, and where the minus (plus) signs correspond to
case that the masing occurs at peri- (apo-) centre. Thus, if circular
orbits are assumed but the orbits have in fact a small ellipticity,
the true distance is smaller (larger) than the inferred distance by
approximately twice the true orbital eccentricity if the systemic
masers occur at peri- (apo-) centre. We have not investigated the case
that the major axis is not aligned with the line of sight, but we
surmise that a case will result intermediate between the peri- and
apo-centre situations, so that the true distance will be uncertain by
a factor of approximately $(1 \pm 2\,e)$.
\vspace*{-1.5em}
\subsubsection{Light Echo Methods}
    \label{sec:Light_Echo_Methods}
``Light Echo'' methods have been proposed and applied to the
absorption of the super nova (SN) ``pulse'' by gas surrounding the
progenitor, which is then re-emitted in the form of line
radiation. This method has been applied to SN 1987A in the LMC by
several authors [e.g., \citet{Pea1991, G1995,GU1998, P1999}], but it
leads to a systematic (unresolvable) distance uncertainty of about
10\% \citep{AG2000}. So far, this method has only been applied to
SN~1987A.

Alternatively, one can attempt to use the polarized reflection of the
SN pulse off dust in the curcum-stellar material. The polarization
condition strongly favors scattering by 90$^o$, so that the physical
radius equals the speed of light times the time-lag between the SN
pulse and the detection of the polarized scattering event
\citep{bS1994}. If the proper motion of the light echo can be
determined, the dependence on the assumed scattering angle is largely
eliminated \citep{bS1996}. Although previous searches for such events
have not produced the hoped-for results \citep{BSM1999, Rea2005_LE},
current workers indicate that the method works quite well (Sugerman,
2006, private communications).

A variant of the ``pulse'' method has recently been developed by
\citet{DB2004}. Their method employs strong X-ray
sources behind nearby galaxies. The dust in the foreground galaxy
scatters the X-rays to produce a halo around the source position. If
the background source is variable, then intensity of this halo will
vary with time and position, where the time-lag ($\Delta t$) is
directly related to the angular distance ($\theta$) from the source
and distance ($D$) to the foreground galaxy: $\Delta t \apx 140 \, D
\, \theta_{100}^2$ days, with $D$ in Mpc and $\theta_{100}$ in units of
100 arcsec. For M~31 \citet{DB2004} estimate that a Chandra observing
campaign of 5C~3.76 with a duration of 2 months (4 months) would yield
a distance accuracy of 4\% (1\%).  However, less observing time would
be required if observing starts after an ``outburst'' of the
background source. This method might also be applied if a gamma-ray
burst would occur behind a nearby galaxy.

The ``Expanding Photosphere'' method can also be used to model the
light curve and expansion history of supernovas. In this method the
distance follows from the measured angular size and the physical size
calculated from integration of the radial velocities.
\citet{Mea2002_SEAM} developed the ``Spectral-fitting Expanding
Atmosphere Method'' which is specifically tailored to the time-varying
and non-LTE nature of SN atmospheres (ejecta). This method has
distance accuracies of order 10\% and has been applied out to z=0.3
\citep{Nea2006_SEAM}.
\subsubsection{Binary Star Methods}
    \label{sec:Binary_Star_Methods}
Detached Eclipsing Binaries (DEBs) are expected to be good distance
indicators throughout the Local Group, possibly as good as 1\%
\citep{PaS1997,P1997,Fea2003,W2004}. In this method, photometric
observations of the eclipses determine the period, the flux ratio and
the ratio of the stellar radii ($R_*$) and the size of the orbit,
while radial velocity observations determine the latter quantity
directly. Spectroscopy and/or multi-band photometry can yield the
effective temperature ($T_{eff}$) of the stars. The absolute
luminosity ($L_*$) can be determined if the relation between surface
brightness ($F_*$) and $T_{eff}$ is reliably determined, so that: $L_*
= 4 \pi R^2 \, F_*$. After correction for extinction, the observed
brightness ($F_{obs}$) then yields the distance: $d = \sqrt{ L_* / ( 4
\pi F_{obs} )} = R_* \, \sqrt{ F_* / F_{obs} }$. After 10 years of
active research [e.g., \citet{Gea1998, Kea1998, Sea1998, Rea2002,
Fea2003, Rea2004, Rea2005, Hea2005_DEBs, Bea2006}], the best
accuracies are currently about 5\%. Possible reasons for the accuracy
gap are the difficulty to determine the effective temperature, surface
brightness and extinction. Also, the early-type main-sequence stars
that are used in the DEB method have no local analogs with accurate
distances and surface brightness measurements: none of the 150
\HIPPARCOS\ stars earlier than B3 have a parallax better than 5\%. We
expect that the \GAIA\ data will provide the necessary calibration
data (distances and extinction) to fully exploit the distance-scale
potential of DEBs. Finally, DEBs are not ideal distance indicators:
when using semi-detached systems or over-contact systems there are
fewer (stellar) parameters to be solved for so that more accurate
distances can be obtained [Wilson, 2002, private communications;
\citet{WW2002, W2004}].

A fundamental-physics method involving binary stars is via the
detection of gravitational waves of close white dwarf binary (CWDB)
stars with LISA. After 10 years of LISA operation, \citet{CS2005}
expect to be able to obtain a distance error for the LMC of order
$4.5\% \times \sqrt{N_{CWDB}/22}$, where the expected number of CWDBs
($N_{CWDB;exp}=22$) may be uncertain by a factor of ten.
  \section{Implications for Cosmology}
\label{sec:Implications_for_Cosmology}
There are at least four reasons to determine $H_0$ with an {\em
accuracy} of several percent or better: 1) to determine distances to
external galaxies via redshift measurements, 2) to calibrate
accurately other distance indicators, 3) to determine the equation of
state of Dark Energy, and 4) to determine the total density
($\Omega_{tot}$) of the universe. The equation of state (EOS) of Dark
Energy (DE) tells us something about its nature.

Several lines of evidence suggest that we live in a Universe that is
close to critical density ($\Omega_{tot} \sim 1$). For example, the
\WMAP\ data and the \HST\ constraint on the Hubble constant ($h = 0.74
\pm 0.08$) results in $\Omega_{tot} \sim 0.996$ and a mass density
(baryons \& dark matter) of 23\%, so that about 76\% of the
energy-density of the universe is unrelated to gravitating matter. The
most obvious candidate for this Dark Energy is a cosmological constant
$\Lambda$, which is allowed for by General Relativity. Other forms of
DE are suggested by theoretical physics. One difference between these
various possibilities is the equation of state of the proposed DE
candidates. For example, the cosmological constant would have $w =
-1$, cosmic strings have $w = -1/3$, domain walls have $w = -2/3$,
while Quintessence can come in multiple varieties: fixed with $w \ga
-0.8$, or with a time-variable $w$ value \citep{PR2003}.  We discuss
two scenarios. In \S\S\ref{sec:A_Flat_Universe} and
\S\S\ref{sec:Including_Curvature} we discuss the case of a constant
$w$, while in \S\S\ref{sec:Comparison_with_the_results_from_the_DETF}
we discuss the case of varying $w$, and compare our results with the
recent conclusions of the Dark Energy Task Force \citep{DETF2006}.

Hu (2005; hereafter H05) analyzes in quite some detail how
measurements of $H_0$ and its variation with redshift would be
beneficial for our understanding of dark energy and its evolution. He
concludes that: ``... the Hubble constant is the single most useful
complement to CMB parameters for dark energy studies ... [if $H_0$ is]
... accurate to the percent level ... .''  However, Hu does not
explicitly describes how errors on the dark-energy parameters are
related to errors on $H_0$, while such information is of pre-eminent
importance when justifying and planning observational campaigns to
determine $H_0$, especially if space missions are considered. In the
remainder of this section, we present some analytical estimates for
the relationship between the errors on $H_0$ and the dark-energy
parameters. These relations are approximate and are intended to
facilitate estimating the required observing time (and cost) of
observing programs aimed at determining $H_0$.

For the reader's convenience, we here present a short summary of Hu's
(2005) description as to how CMB data physically relate to the Hubble
constant. In fact, the CMB experiments determine the photon-to-baryon
ratio ($R_*$) and the matter-to-radiation ratio ($r_*$) from the
locations and ratios of the acoustic peaks. H05 finds:
$r_* \propto \frac{0.126   }{\omega_m} \frac{1+z_*}{1089}$, and 
$R_* \propto \frac{\omega_b}{0.0223  } \frac{1089}{1+z_*}$,
where $\omega_b \equiv \Omega_b h^2$ is the physical baryon density.
These relations allow us to derive the errors ($\epsilon$) on $R_*$
and $r_*$ from the errors on $\omega_m$ and $\omega_b$ as given by
WMAP07. The results are: $\epsilon_{r_*}$=7.1\% and
$\epsilon_{R_*}$=3.6\% for $\omega_m = 0.126 \pm 0.009$, and $100\,
\omega_b = 2.23 \pm 0.08$.
%
%
Physically, the co-moving size of acoustic oscillations ($s_*$) equals
the distance a sound wave has traveled during the time between the Big
Bang and recombination, and depends on $R_*$, $r_*$, $z_*$ and the
sound speed (and hence the matter density $\omega_m$): $s_* \approx
139.8 \, (\frac{R_*}{0.854})^{-0.252} \, (\frac{r_*}{0.338})^{0.083}$
Mpc (scaled from H05), with a 1\% uncertainty. Note that $z_*$ depends
only weakly on cosmology. The CMB data also yields the location of the
acoustic peak ($\ell_A$) with smaller error than that of $s_*$. Thus,
the angular-diameter distance of the acoustic peak ($D_A$) is
well-established observationally: $D_A \equiv \ell_A s_* / \pi$, while
theoretically $D_A$ depends on cosmology:
\begin{eqnarray}
D_A(\Omega_K=0)  &=&  a_* \int_{a_*}^1 \frac{1}{a^2 H(a)} \, \, da
   \label{eqn:D_A}\\
H(a) &=& H_0
     \sqrt{ 
        \frac{\Omega_{m}      }{a^{3     }} + 
        \frac{\Omega_{\Lambda}}{a^{3(1+w)}}
          }
   \label{eqn:H_a}
\end{eqnarray}
where we neglect the small contribution from the density in
relativistic particles ($\Omega_\nu/a^4$), and where $a=1/(1+z)$ is
the scale factor of the universe. If the curvature density
($\Omega_K$) does not vanish, eqn.~(\ref{eqn:D_A}) needs to be
modified [e.g., \citet{C2001}].  In this simple model, the sum of the
$\Omega$-terms equals unity so that:
\begin{eqnarray}
\Omega_\Lambda &\equiv& 1 - \Omega_m \, = \, 1 - \omega_m / h^2
   \approx 0.770 \pm 0.048 \, \, ,
   \label{eqn:Omega_L_Omega_m_OT1}
\end{eqnarray}
where the approximate value for eqn.~(\ref{eqn:Omega_L_Omega_m_OT1})
follows from the CMB value for $\omega_m$ and the \HST\ value for $h$.
Thus, accurate determinations of $\omega_m$ and $h$ would
significantly constrain the dark-energy. However, because the measured
$D_A$ is an integral constraint, it reveals nothing about the
variation of $H$ with scale-factor (or redshift or time). If in
reality $w$ varies with scale factor, while it is assumed to be
constant, then application of the integral constraint would result in
an erroneous value of $H_0$ and/or $w$.  However, if $w$ is constant,
as would be the case if the DE results from the cosmological constant,
cosmic string, domain walls, etc., then a determination of $H_0$ would
nail down $w$.  Because the variation of $w$ is an important
discriminator between various possible dark-energy candidates [see,
e.g., \citet{PR2003, F2006, AMNO2006, DETF2006}, and references
therein] one would want to determine the variation of $w$ accurately.

\subsection{A Flat Universe}
 \label{sec:A_Flat_Universe}
Here we consider the case that the universe is flat, and if the
equation of state is constant. For this simple case, we can determine
the EOS of dark energy from the CBM data and the Hubble constant
only. We use this example to illustrate that dark-energy information is
contained in accurate knowledge of $H_0$.  For the case of a flat
universe ($\Omega_K=0$) and with $\Omega_\nu=0$, eqn.~(\ref{eqn:H_a})
can be simplified to read:
\begin{eqnarray}
\frac{\overline{H}(a)}{100} &=&
     \sqrt{ \frac{\omega_{m}    }{a^{3     }} + 
            \frac{h^2 - \omega_m}{a^{3(1+w)}} }
   \label{eqn:Hb_a}
\end{eqnarray}
If the dark energy is in the form of a cosmological constant with
$w=-1$, then the only unknown left in eqn.~(\ref{eqn:Hb_a}) is the
Hubble constant which can be found from the constraint provided by the
observed value of $D_A$ [eqn.~(\ref{eqn:D_A})]. From $h$, $\Omega_m$
and $\Omega_\Lambda$ follow trivially. This is the basis for the
reported determination of the Hubble constant from the \WMAP\
data. However, the inferred value of the Hubble constant (and hence
$\Omega_m$ and $\Omega_\Lambda$) depends on the choice of the EOS of
the dark energy [via the $1/a^{3(1+w)}$-term in
eqn.~(\ref{eqn:Hb_a})]. This relationship is illustrated in Fig.~15 of
WMAP07, from which we derive:
\begin{eqnarray}
   -w \hspace*{-0.7em} &\approx& \hspace*{-0.8em} 1.59 - 2.56 \, \Omega_m 
         =    1.59 - 2.56 \, \frac{\omega_m}{h^2 }
   \label{eqn:w_omega_h}
\end{eqnarray}
where the second equality is re-written in terms of the observables.
The errors on the EOS and the density of dark energy for the case of a
flat universe are thus:
\begin{eqnarray}
   \epsilon_{\Omega_\Lambda} &=& \phantom{2.56 \,} \Omega_m \, \sqrt{
      \left( \frac{  \epsilon_{\omega_m}}{\omega_m} \right)^2 +
      \left( \frac{2 \epsilon_{h       }}{h       } \right)^2 } 
      \, \approx \, 6.3\%
   \label{eqn:Delta_Omega_L}\\
  \epsilon_w &\approx& 2.56 \, \Omega_m \, \sqrt{
      \left( \frac{  \epsilon_{\omega_m}}{\omega_m} \right)^2 +
      \left( \frac{2 \epsilon_{h       }}{h       } \right)^2 }
      \, \approx \, 12.3\%  \,\, .
  \label{eqn:Delta_w_omega_h}
\end{eqnarray}
With uncertainties of 7.1\% for $\omega_m$ and 9.8\% for $h$
(including the systematic error term) the uncertainties on $w$ and
$\Omega_\Lambda$ are almost evenly divided between the two error
terms. However, in the future this will no longer be the case because
the \PLANCK\ mission is expected to reduce the error on $\omega_m$ by
a factor of eight \citep{PLANCK}. This indicates that in the
post-\PLANCK\ era, our knowledge of the value of dark energy will be
fully dominated by the uncertainty on $H_0$ {\em if} the (systematic)
error on $H_0$ is not decreased significantly, and if no other data is
considered. \citet{BABR2004} reach a similar conclusion albeit that
they argue that the CMB data imparts a correlation between $\omega_b$
and $\omega_m$ that will yield the same $s_*$ so that it is impossible
to go below $\epsilon_w \sim 0.1$ employing $H_0$ and CMB data
alone. We note that this assertion is at odds with H05's who concludes
that $\omega_b$ and $\omega_m$ are virtually independently determined
from the CMD data.

\subsection{Including Curvature}
 \label{sec:Including_Curvature}
When adding other relevant data sets, the various cosmological
parameters are determine better than with \WMAP\ data alone.
Inspecting Figure 21, Table 11 and Figure 17 of WMAP07, we derive the
following $\chi$-by-eye relations for $\Omega_\Lambda$, $\Omega_K$ and
$w$, respectively:
\begin{eqnarray}
   \Omega_\Lambda &=& a_{\Lambda m} + b_{\Lambda m} \Omega_m
   \label{eqn:Omega_Lm} \\
   \Omega_K &=& a_{K\Lambda} + b_{K\Lambda} \Omega_\Lambda
   \label{eqn:Omega_KL} \\
   w        &=& a_{wK} + b_{wK} \Omega_K \, \, .
   \label{eqn:w_K}
\end{eqnarray}
We estimate the coefficients to equal: $a_{\Lambda m} \sim 0.944 \pm\
0.011$, $b_{\Lambda m}\sim -0.775$ and $a_{K\Lambda} \sim -0.0992 \pm\
0.009$, $b_{K\Lambda} \sim 0.1199 \pm 0.0124$ (the $K\Lambda$-terms
are determined excluding the $H_0$ constraint in Table 11 of WMAP07)
and $a_{wK} \sim -0.910 \pm 0.063$, $b_{wK} \sim 6$. In order to
investigate the effects of the Hubble constant, we combine equations
(\ref{eqn:Omega_Lm}) through (\ref{eqn:w_K}) in terms of $h$ and
arrive at:
\begin{eqnarray}
&& \hspace*{-2.8em}
   \Omega_\Lambda = (0.944 \pm\ 0.011) -0.775 \frac{\omega_m}{h^2}
   \label{eqn:Omega_Lambda_General}\\
   &\approx& 0.766 \pm 0.020 \,\, (4.9\%)
   \nonumber\\
&& \hspace*{-2.8em}
   w = a_{wK}      + a_{K\Lambda} b_{wK} +
       a_{\Lambda m} b_{K\Lambda} b_{wK} + 
       b_{\Lambda m} b_{K\Lambda} b_{wK}  \frac{\omega_m}{h^2}
   \label{eqn:w_theory}\\
     &=& (-0.826 \pm\ 0.109) - (0.557 \pm\ 0.058) \, \frac{\omega_m}{h^2}
   \label{eqn:w_now_rel} \\
     &\approx& -0.95 \pm\ 0.11 \,\, (11.6\%) \, ,
   \nonumber
\end{eqnarray}
where the relations that are not numbered employ the values and errors
on $\omega_m$ from the CMB and $h$ from \HST. Comparing the relation
for $w$ for a flat universe [eqn.~(\ref{eqn:w_omega_h})] with the
general case [eqn.~(\ref{eqn:w_now_rel})] we notice two things. First,
the error is larger for the general case, presumably due to the extra
degree of freedom that arises from fitting for $\Omega_K$. Second, the
dependency on the value of $H_0$ has decreased by a factor of almost
five as evidenced by the smaller coefficient of the term containing
$h$. This is probably due to the fact that the relations for the
general case were derived employing the other data sets. Also note
that $\Omega_\Lambda^{flat}$ equals $\Omega_\Lambda^{general}$ to
within the errors, and that the derived errors are approximately the
same for the two cases.

For the general case, we can now also obtain an expression for the
total density:
\begin{eqnarray}
   \Omega_{noK} &=& \Omega_\Lambda + \Omega_m \, = \, a_{\Lambda m} + 
                  ( b_{\Lambda m} + 1 ) \, \frac{\omega_m}{h^2}
   \label{eqn:Omega_tot}\\
    &=&    (0.9438 \pm\ 0.0114) + 0.225 \,\frac{\omega_m}{h^2} 
   \label{eqn:Omega_now_rel} \\
    &\approx& 0.996 \pm\ 0.016 \,\, (1.6\%) \, .
   \label{eqn:Omega_now_val}
\end{eqnarray}
Thus, without the assumption of a flat Universe, current data allows
for the determination of the total density of the universe to plus or
minus 1.6\%, while the EOS of Dark Energy is known to about
12\%. The actual values of $\Omega_{tot}$ and $w$ strongly hint at a
flat Universe with the cosmological constant being the Dark Energy.

\subsection{$H_0$ and Dark Energy}
   \label{sec:H_0_and_Dark_Energy}
Future CMB data will be more accurate than at present: the eight-year
\WMAP\ data will reduce the current errors by a factor of two (Spergel,
2006, private communications), while \PLANCK\ data is expected to be
eight times better than the 3-year \WMAP\ data \citep{PLANCK}.

We can estimate the effects of more accurate CMD data, as well as an
accurate value of the Hubble constant on our knowledge of the EOS of
Dark Energy by employing equation (\ref{eqn:w_theory}) above. In order
to do so, we assume that the errors on the $a_i$ and $b_i$ parameters
improve while the values themselves do not change with improved CMB
accuracy. The advantage of this method is that it is easily
implemented. A full, WMAP07-style, calculation could be performed, but
is beyond the scope of this paper.

As discussed in \S\ref{sec:Implications_for_Cosmology}, the physics of
the CMB implies that only a {\em relation} can be found between
$\Omega_\Lambda$ and $\Omega_m$. This leads to a very elongated
confidence region. Other data sets are required to break this
degeneracy. For example, the magnitude-redshift relation for
Supernova~Ia is popular because its confidence region is almost
perpendicular to that of the CMB data [for reviews, see among others,
\citet{C2001, LB2002, Rea_SN2004, PLANCK, P2005}]. In that case, the
SN-Ia data ``selects'' part of the confidence region generated by the
CMB data to determine {\em values} for $\Omega_\Lambda$ and
$\Omega_m$. However, it will hardly decrease the {\em error} in
$\Omega_\Lambda$ (for a given $\Omega_m$) because the SN-Ia confidence
region is very elongated in the $\Omega_\Lambda$ direction.  Thus,
orthogonal constraints are the way to determine parameter {\em
values}, while the shape and direction of the joint confidence region
is set by the data set with the smallest errors. In the remainder of
this paper we assume that the CMB measurements will be the most
accurate data set, so that the relations between the cosmological
parameters given by equations (\ref{eqn:Omega_Lm}) through
(\ref{eqn:w_theory}) and (\ref{eqn:Omega_tot}) are approximately
valid. We will also investigate the consequences of relaxing this
assumption.

\begin{figure}
   \hspace{-0.50em}
      \includegraphics[width=85mm,height=110mm]{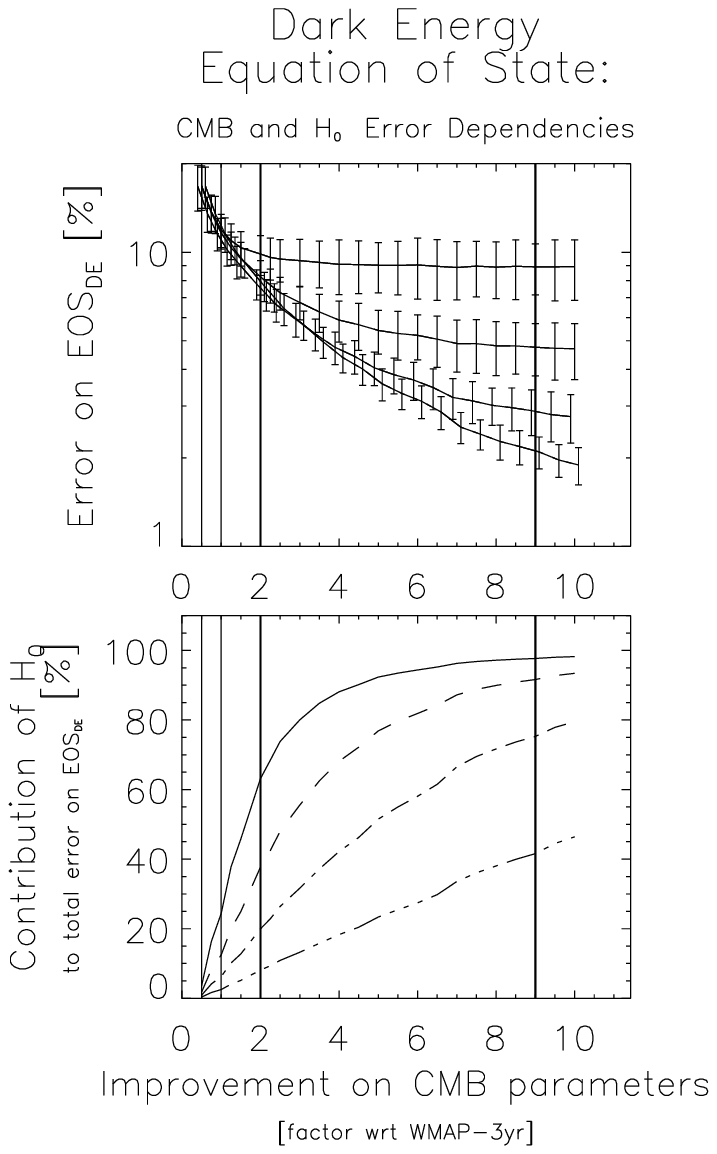}
   \caption{ \label{fig:Cosmology} {\em Top Panel:} The accuracy with
   which the EOS of Dark Energy can be determined as a function of the
   accuracy of the CMB data (abscissa), and the accuracy in $H_0$
   (curved lines). The vertical lines correspond to accuracies of the
   \WMAP-1-year data (left), \WMAP-3yr data, \WMAP-8yr data and
   \PLANCK\ (right). With \PLANCK-like CMB parameters and $H_0$
   accuracies of, from top to bottom, 10.8\% (8 \kmsMpc), 5.4\% (4
   \kmsMpc), 2.7\% (2 \kmsMpc) and 1.1\% (0.8 \kmsMpc), the achievable
   errors on $w$ are: 8.9\%, 4.8\%, 3.0\% and 2.3\%, respectively. The
   ``error bars'' represent the case that the slopes $b_i$ in
   eqns. (\ref{eqn:Omega_Lm}) -- (\ref{eqn:w_K}) are randomly varied
   by \pmt 25\%. The lines in the {\em bottom panel} show the
   contribution of the uncertainty on $H_0$ to the total error on $w$
   for $\epsilon_{H_0}$=8/1, 8/2, 8/4 and 8/10 \kmsMpc, from top to
   bottom.}
   \vspace{-1em}
\end{figure}

The results are presented in figure~\ref{fig:Cosmology}, where the top
panel shows $\epsilon_w$ as a function of improvement of our knowledge
of the CMB parameters, with respect to the \WMAP\ 3-year data. The
four lines are computed for Hubble constants with errors of 10.8\%,
5.4\%, 2.7\% and 1.1\%, from top to bottom. These errors correspond to
improvements with respect to the current error by factors of 1, 2, 4
and 10, respectively. The corresponding errors on $w$ that are
attainable with \PLANCK-like data are, 8.9\%, 4.8\%, 3.0\% and 2.3\%,
respectively.  These errors are computed via the standard error
propagation relations for equation (\ref{eqn:w_theory}). The part of
the error on $w$ that can be attributed to uncertainty in $H_0$
($\epsilon_{w,h}$) depends on the relative accuracy of the CMB
parameters and $H_0$.  Currently, the error on $H_0$ hardly affects
our knowledge of $w$. However, with decreasing errors on the CMB
parameters, our ignorance of the Hubble constant becomes the dominant
contribution to the total error on $w$ ($\epsilon_{w,tot}$). This is
illustrated in the bottom panel of figure~\ref{fig:Cosmology} which
shows $\epsilon_{w,h} / \epsilon_{w,tot}$ as a function of the
accuracy of CMB data and the Hubble constant.

From figure~\ref{fig:Cosmology} we infer that at \PLANCK\ accuracy,
the errors on $w$ will have only slightly decreased with respect to
the current value of 11.5\%. However, the accuracy on the EOS of Dark
Energy improves significantly when the error on $H_0$ is decreased
significantly. In section
~\ref{sec:Comparison_with_the_results_from_the_DETF} below we estimate
the effects of significant improvements in the quality of the
other data sets and show that in those cases the effects of
improved knowledge of $H_0$ diminishes.

\subsection{$H_0$ and $\Omega_{tot}$}
   \label{sec:H_0_and_Omega_tot}
The deviation from spatial flatness [$(1-\Omega_{tot}) \equiv
\Omega_K$] follows from Equation~(\ref{eqn:Omega_now_val}):
$\Omega_K \sim 0.004 \pm 0.016$.  Thus, the spatial curvature is
currently known to within a factor of four. We applied the same
error-propagation methods for $\Omega_{tot}$ as for $w$ and find a
behavior for the error on $\Omega_{tot}$ very similar to $\epsilon_w$,
but at approximately ten times lower level. Thus, a determination of
the Hubble constant at the 1\% level would decrease the error on the
spatial curvature by a factor 16: from a factor four to 25\%.
\subsection{Comparison with the results from the Dark Energy Task Force}
 \label{sec:Comparison_with_the_results_from_the_DETF}
In their recent report, the Dark Energy Task Force [DETF;
\citet{DETF2006}] presented recommendations as to how to increase our
knowledge of dark energy, and in particular its evolution. As a
starting point, they used the CMD data as expected to be delivered by
\PLANCK\ and added to that the expected improvements of a large
variety of other proposed space-based and/or ground based
investigations based on Baryon Acoustic Oscillations, Galaxy Clusters,
Supernovae and Weak Lensing. The uncertainty on $H_0$ used by the DETF
corresponds to the current systematic $H_0$ error as derived from the
\HST\ data \citep{H0HST}.

Our work presented above is based on the case of a fixed EOS of the
dark energy, where we also assumed that any future increase in
accuracy would be dominated by the \PLANCK\ results, and that the
accuracies of the other data sets would remain the same. In
DETF-speak, we use ``Stage~I'' accuracies for the other data sets and
find $\epsilon_w$=8.9\%. From the models tabulated by the DETF, we
infer $\epsilon_w$ values of approximately 3.6\%, 2.4\% and 1.45\% for
the combined Stage~II, Stage~III and Stage~IV data,
respectively\footnote{
The DETF considered explicitly DE models with varying values of the
EOS: $w(a) = w_0 + (1-a)\, w_a$, with $w_0$ the value at the current
epoch and $w_a$ the slope of the EOS versus scale-factor relation. For
any fitted linear relation, there is point (the ``pivot point'') where
the total error is minimal. In this case, $w_p = w_0 + (1-a_p)\, w_a$,
and $a_p$ scale factor of the pivot point.  The DETF used a figure of
merit ($FoM$) to evaluate the effectiveness of the other data
sets. The $FoM$ is essentially $FoM = 1/(\epsilon_{w_p} \,
\epsilon_{w_a})$, .  For models such as ours that do not include EOS
evolution ($w_a\equiv0$), the ``pivot point'' lies at $a_p=1$. We thus
compare our $\epsilon_w$ with $\epsilon_{w_p}$ as tabulated by the
DETF.
}. But note that these accuracies are only reached if all programs per
stage are combined, and that all the programs in the previous stages
are also executed.

\begin{figure}
   \hspace{-0.50em}
      \includegraphics[width=85mm,height=140mm]{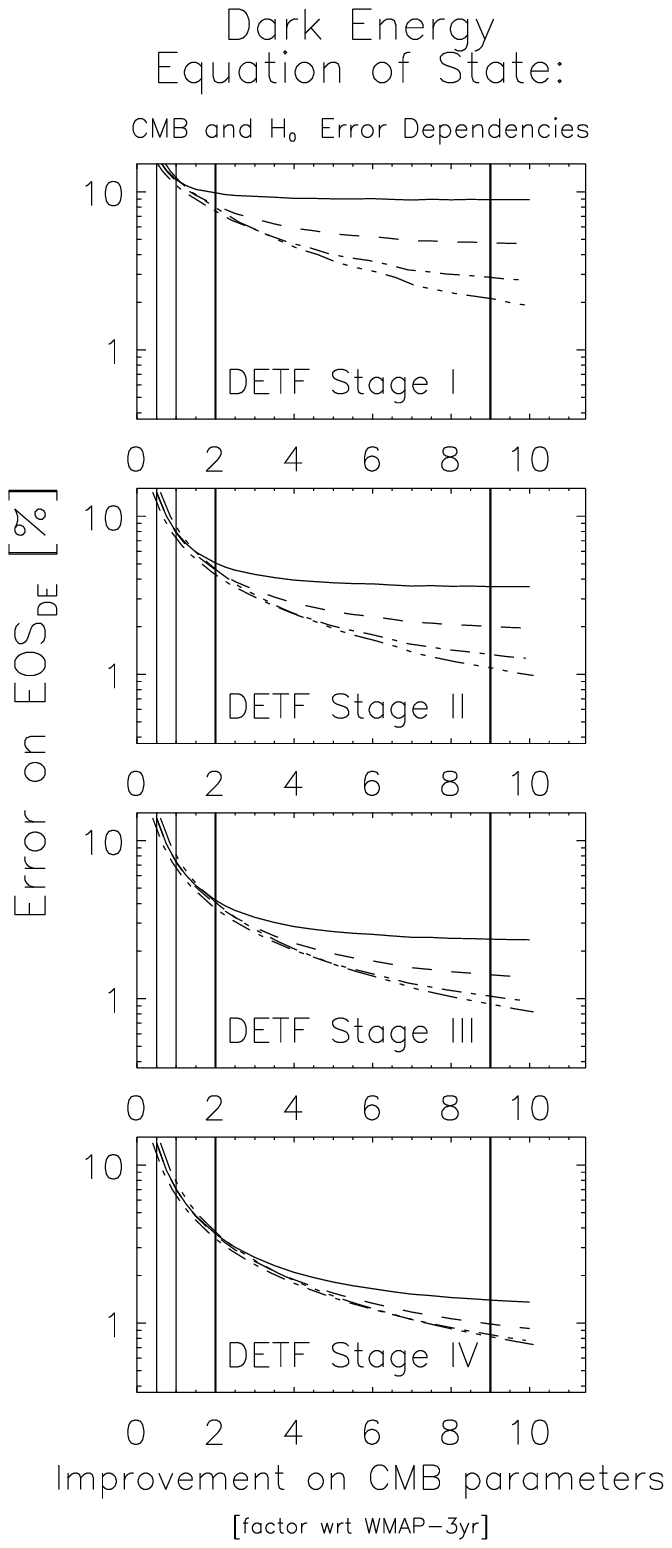}
   \caption{ \label{fig:Cosmology_DETF} In each panel we present the
   accuracy with which the EOS of Dark Energy can be determined as a
   function of the accuracy of the CMB data (abscissa), and the
   accuracy in $H_0$ (curved lines). The vertical lines correspond to
   accuracies of the \WMAP-1-year data (left), \WMAP-3yr data,
   \WMAP-8yr data and \PLANCK\ (right). The difference between the
   four panels is the accuracy with which the other data sets are
   determined. From top to bottom, these accuracies correspond
   approximately to stages I, II, III and IV as defined by the Dark
   Energy Task Force (see text,
   \S\S\ref{sec:Comparison_with_the_results_from_the_DETF}),
   respectively. In each of the panels, the four lines correspond to
   the same errors on $H_0$: $\epsilon_{H_0}$=8/1, 8/2, 8/4 and 8/10
   \kmsMpc, from top to bottom. The attainable errors with \PLANCK\
   data are: $\epsilon_w$= 8.90\%, 4.81\%, 3.02\% and 2.27\% for
   Stage~I (as in Fig.~\ref{fig:Cosmology}; $\epsilon_w$= 3.60\%,
   2.06\%, 1.43\% and 1.20\% for Stage~II; $\epsilon_w$= 2.41\%,
   1.48\%, 1.14\% and 1.02\% for Stage~III; $\epsilon_w$=1.45\%,
   1.07\%, 0.95\% and 0.91\% for Stage~IV.}
   \vspace{-1em}
\end{figure}
Unfortunately, we can not directly compare our results with those of
the DETF, but we can crudely scale our methodology to their results in
the following manner. In our analysis, the strongest correlation
(degeneracy) between parameters occurs for $w$ and $\Omega_K$, with a
coefficient of $b_{wK}\sim 6$ [cf. eqn.~(\ref{eqn:w_K})]. The effect
of adding more accurate data is to break the degeneracies between the
cosmological parameters, which we implement by decreasing the value of
$b_{wK}$. After some experimentation, we find that we recover the DETF
$\epsilon_w$ values if we decrease $b_{wK}$ by factors of 2.51, 3.90
and 7.51 for Stages II, III and IV, respectively. The DETF tested the
effects of smaller errors on $H_0$ for Stage~IV data, and conclude
that a factor of two improvement would lead to a reduction of
$\epsilon_w$ by at most 50\%, while we find an improvement of 26\%,
which is consistent with the DETF results.

Employing this calibration, we estimate the effects of smaller
$\epsilon_h$ values as a function of the accuracy of the other
data. The results for Stages~I, II, III and IV are presented in the
top through bottom panels of Figure~\ref{fig:Cosmology_DETF},
respectively. Note that the top panel corresponds to our results
discussed above (and Fig.~\ref{fig:Cosmology}), while the bottom panel
corresponds to the best-case studied by the DETF. As expected, the
effects of an improved error for $H_0$ has less and less of an effect
on $\epsilon_w$ as the other data becomes more and more accurate.

Figure~\ref{fig:Cosmology_DETF} suggests that some improvement in the
other data sets can be traded for an improved error on $H_0$. That is
to say, Stage~I with $\epsilon_h/20$, Stage~II with $\epsilon_h/4$,
Stage~III with $\epsilon_h/2$ and Stage~IV with $\epsilon_h/1$ yield
similar accuracies for the equation of state of dark energy. [We give
the exact numerical results in the caption of
Figure~\ref{fig:Cosmology_DETF}.] The DETF quotes that {\em each} of
the Stage~III projects would cost tens of millions of dollars (US),
while the Stage~IV projects might be as expensive as 300 -- 1,000 M\$
{\em each}, or a total of at least 2,000 M\$. However, the primary
motivations for each of these Stage~IV projects are not necessarily
related to dark-energy science. Nevertheless, we suggest that
significant cost and time savings might be obtained if more effort is
expended on determining a more accurate $H_0$, or at least that it
would be worthwhile to investigate the matter more thoroughly. For
example, a ``VLBA'' effort to find and analyze more extra-galactic
water masers may reduce $\epsilon_h$ by a factor of two to five at a
cost of maybe several M\$, while a SIM-based rotational parallax
program can reduce $\epsilon_h$ by a factor of 5--10 and might cost 20
M\$. In fact, such a ``VLBA'' effort is already underway
\citep{WMCP2006}.
  \section{Rotational Parallax Distances}
\label{sec:Rotational_Parallax_Distances}
The method of ``Rotational Parallaxes'' (RP) combines proper motions
and radial velocities of stars in external galaxies to yield bias-free
single-step distances, analogous to the orbital parallax
technique. The attainable accuracies, of the order of one percent, are
well matched to the requirement that the interpretation of future CMB
data is not limited by knowledge on $H_0$.

For a nearby spiral galaxy at distance $D$ (in Mpc) which is inclined
by $i$ degrees, with a rotation speed of $V_c$ \kms, the proper motion
due to rotation equals $\mu_{RC} = \frac{V_c}{\kappa D}$ \muasyr,
where $\kappa \approx 4.74$ [\kms]/[AU yr\rtp{-1}]. For M\,33 [$i \sim
56^o, D \sim 0.84, V_c \sim 97$], M\,31 [$i \sim 77^o, D \sim 0.77,
V_c \sim 270$] and the LMC [$i \sim 35^o, D \sim 0.055, V_c \sim 50$]
we find: $\mu_{RC}^{M33} \sim 24$, $\mu_{RC}^{M31} \sim 74$ and
$\mu_{RC}^{LMC} \sim 192$ \muasyr. Thus, the rotational motions of
these galaxies are easily resolved by \SIM, and, depending on
magnitude, also by \GAIA\ and \OBSS\footnote{
\label{foot:astrometric_errors} We performed extensive simulations to
convert accuracies in position or parallax to proper motion errors,
with as a variable the inverse of the mission duration ($x_t$). We
find that $\epsilon_\mu / \epsilon_{pos} \approx 1.901 \, x_t + 0.474 \,
x_t^2$, while $\epsilon_\mu / \epsilon_\pi \approx 2.751 \, x_t + 0.981 \,
x_t^2$. Thus, for mission durations of 5 years, a position error at
V=18 of 5.38 \muas\ for \SIM\ correspond to proper motion errors of 2.1
\muasyr. \SIM's best grid accuracy is 3.47 \muas\ at V=11, so that
$\epsilon_{\mu_\sim}$ 1.35 \muasyr. We estimate, based on \GAIA's scanning
law, read noise, collecting area, quantum efficiency and integration
time, that it can achieve parallax Eros of 26, 45 and 78 \muas\ at
V=15, 16 and 17, respectively. This leads to proper motion errors of
15, 26 and 46 \muasyr, at the same magnitudes.
%
%
}. In addition to the absolute size of the rotation speed, the
relative contribution of random motions of the stellar population
($\sigma$) are important. The $V_C/\sigma$ values of 27, 9.7 and 2.5
for M\,31, M\,33 and the LMC are likely to be indicative of the ease
with which accurate RP distances can be determined.

The ``principal axes'' (or $Mm$) variant of the RP method applies to
galaxies dominated by circular rotation.  Individual stars are
identified along the major ($M$) and minor ($m$) axes at similar
distances from the centre (with the same rotational velocity. Proper
motions ($\mu$) on the minor axis (\mumC) measure the circular
velocity divided by $D$. On the major axis, \muMC, equals \mumC\
projected by $\cos i$. The correction for the systemic motion
(indicated by superscript ``$C$'') can be approximately achieved when
{\em two} stars are chosen per principle axis, one on each side of the
centre (e.g., $\abs{\muMC} \, = \, \abs{\mu_{M,approaching} -
\mu_{M,receeding}} / 2$).  Radial velocities (\Vr) on the major axis
give the circular velocity projected by $\sin i$.  Combining these
yields \citep{PS1997}:
\begin{eqnarray}
\cos i   &=&   \frac{\abs{\muMC}}{\abs{\mumC}}
   \label{eqn:cos_i_Mm} \\
 V_c     &=& \frac{\Vr \, \mumC}{\sqrt{\mumCsq - \muMCsq}}
   \label{eqn:V_c_Mm} \\
D_{Mm}    &=& \frac{\Vr}{\sqrt{\mumCsq-\muMCsq}} \, .
   \label{eqn:D_Mm}
\end{eqnarray}
However, the principal axes method requires stars close to the
principal axes, making it difficult to find enough targets, while the
effects of major perturbations such as warps, spiral arms etc. are
difficult to handle.

This method can be generalized to a star arbitrarily positioned in the
galaxy. We use rectangular ($x$ and $y$) and polar ($R$ and $\theta$)
coordinate systems centred on the galaxy centre and co-planer with the
galaxy disk.  The $x$ and $y'$ axes along the major and minor axes,
respectively, with $y'$ is the foreshortened $y$ coordinate.  The
following elementary relations between the coordinates and the various
projections of the orbital velocity $\overline{V_c}$ apply:
\begin{eqnarray}
   \Vx &=& -s_\Omega V_c \sin\theta' \label{eqn:V_x} \,
   \label{eqn:Vx} \\
   \Vy &=&  \ph{-}s_\Omega V_c \cos\theta' \, = \, 
      \frac{-\Vx}{\tan\theta'}
      \label{eqn:V_y} \\
   \Vr &=&  \ph{-s_\Omega} \Vy \sin i\ph{'}
       \,\ =  \, s_\Omega V_c \cos\theta' \sin i 
   \label{eqn:V_r} \\
   \Myp &=& \frac{\Vyp}{\kappa \, D}  \, =  \, \frac{\Vy \cos i}{\kappa \, D}
        =  \frac{s_\Omega V_c \cos\theta' \cos i}{\kappa \, D}
   \label{eqn:Myp}\\
   \Mx  &=&  \frac{\Vx}{\kappa \, D} =
        \frac{-s_\Omega V_c \sin\theta'}{\kappa \, D} \, ,
      \label{eqn:Mx}
\end{eqnarray}
with \Vx\ and \Vy\ the projections of the $\overline{V_c}$ vector on
the $x$ and $y'$ axes. The angle $\theta'$ is the angle between
$\overline{V}_c$ and \Vy, while $s_\Omega$ = -1 (+1) for (counter-)
clockwise rotation.

The three unknowns $D, i$ and $V_c$ can be recovered from the three
observables ($\Vr$, $\mu_x$, and $\mu_{y'}$) by solving
eqns.~(\ref{eqn:V_r}) -- (\ref{eqn:Mx}).  Equations~(\ref{eqn:D_S}) --
(\ref{eqn:Vc_Vr_obs}) below codify the ``individual star'' ($IS$)
method:
\vspace*{-0.7em}
\begin{eqnarray}
D_{IS} &=& \frac{\Vr}{\kappa} \, \sqrt{ \frac{-y'/\Myp}{x\Mx+y' \Myp} }
   \label{eqn:D_S}\\
\cos^2 i_{IS} &=& \frac{\phantom{-}y' \, \Myp}{-x \, \Mx}
   \label{eqn:cos2_is}\\
V_{c,IS} &=& \Vr 
        \sqrt{ \frac{\Mx}{\Myp} \, \, 
               \frac{x\Myp - y'\Mx}{x\Mx + y'\Myp} } \, ,
   \label{eqn:Vc_Vr_obs}
\end{eqnarray}
where we assume circular orbits. 

The achievable distance error depends on distance, inclination,
position angle and (almost) linearly on observing errors
(OP2000). Evaluating eqn.~(\ref{eqn:D_S}) at various positions
$(x,y')$ in the target galaxy, and if we assume that the accuracies
for the radial velocity observations ($\epsilon_{V_r}$) and proper
motions are of order the internal velocity dispersion of the tracer
population, we obtain a distance error (per star) of 13\% for
M\,31. For individual stars in M\,33 and the LMC, we get errors per
star of 28\% and 90\%. Note that for M\,31 and M\,33 the random
motions of a young population (10 \kms) corresponds roughly to the
expected proper motion accuracy of \SIM\ of 2.1 \muasyr but is much
smaller than \GAIA's and \OBSS's accuracies.  For the LMC, the
internal motions roughly equal the astrometric errors of \GAIA\ and
\OBSS\footnote{%
For M\,31 and $\sigma
\sim$10 \kms\ $\sim \epsilon_{\Vr}$, and $\epsilon_{\mu} \sim
\sigma/(\kappa 770$ kpc$) \sim 2.7 \ \muasyr 
\sim \epsilon_{\mu_{SIM}} \ \ll \ \epsilon_{\mu_{GAIA}}$, \\
-- For M\,33 and $\sigma \sim$10 \kms\ $\sim \epsilon_{\Vr}$, and
$\epsilon_{\mu} \sim \sigma/(\kappa 840$ kpc$) \sim 2.5 \ \muasyr \sim
\epsilon_{\mu_{SIM}} \ \ll \ \epsilon_{\mu_{GAIA}}$, \\
-- For the LMC, the ``errors'' are dominated by the velocity
dispersion ($\sigma \sim 20$ \kms) of the stars: $\epsilon_\mu \sim
\sigma /(\kappa \, 55$ kpc$) \sim 77 \ \muasyr \gg
\epsilon_{\mu_{SIM}} \ \sim \epsilon_{\mu_{GAIA}} $ }.
Also note that the distance error of 28\% for M\,33 is quite close to
the value obtained from H$_2$O-masers method by \citet{Bea_H2O_2005}.

Thus, for RP programs targeted at either M\,33 or M\,31, the main
obstacle in achieving a small distance error is the smallness of the
rotational signal. \SIM, with its small errors, is thus much better
suited for such a project. On the other hand, \SIM's small errors would
be ``wasted'' on the LMC stars because there the error budget is
dominated by the dispersion of the stellar population.
\subsection{Realistic Rotational Parallaxes}
 \label{sec:Realistic_Rotational_Parallaxes}
If not modeled carefully, non-circular motions due to spiral-arm
streaming motions, perturbations from nearby galaxies, a bar, warps,
etc. may result in a biased distance determination (OP2000).  We have
already seen that the obtainable errors for the idealized case of
stellar-velocity-dispersion effects only yields errors of 13\%, 28\%
and 90\% per star for M\,31, M\,33 and the LMC, respectively. Since
the errors due to random motions are stochastic, these errors would go
down with the inverse of the square-root of the number of ``stars'' in
the sample.  In addition to the uncertainty due to velocity
dispersion, spiral-arm streaming motions can reach amplitudes of 10\%
of the rotation velocity, leading to {\em systematic} distance errors
of similar magnitude. For the LMC the effects of the bar and tidal
interaction are even more important than for M\,31 and M\,33.  In
order to achieve errors of several percent, it will be necessary to
correct for any sizable deviations from circular motion. OP2000
indicate that such can indeed be achieved with the next-generation
astrometric satellites.
\subsubsection{Rotational Parallax: General Case}
    \label{sec:Rotational_Parallax:_General_Case}
For the general case, we separate the three-dimensional motion of a
star in four parts: 1) the systemic motion ($\overline{V}_{sys}$) of
the galaxy, 2) a random velocity $\overline{V}_\sigma$, 3) the
circularly symmetric part of the orbital velocity $\overline{V}_c$,
and 4) the peculiar component $\overline{V}_p$. The idea behind this
separation is to group physical effects together that cause
large-scale correlations between stellar motions that are amenable to
modeling. The systemic and circular terms are easy to model, while the
random term can only be taken into account in a statistical sense. The
ease with which the peculiar velocity can be modeled depends on its
cause. Thus, the total space motion of a star in an external galaxy is
given by:
\begin{eqnarray}
\overline{V}_{total} &=& 
   \overline{V}_{sys} +
   \overline{V}_{orbit}    +
   \overline{V}_\sigma \, , \, {\rm and \, where}
   \label{eqn:V_total} \\
   \overline{V}_{orbit} &\equiv& 
      \overline{V}_{c\phantom{ys}} + \overline{V}_{p} \, .
   \label{eqn:V_vector}
\end{eqnarray}
Each term projects onto the orthogonal directions
$(\hat{x},\hat{y},\hat{z})$. $\overline{V}_{c}$ corresponds to the
rotation speed $V_c$, with components $V_{c,x}=V_c \sin \theta$ and
$V_{c,y}=V_c \cos \theta$, and $\tan\theta = y/x$. If the systemic
motion makes an angle $i_s$ with respect to the sky, the three
observable velocities are:
\begin{eqnarray}
\kappa D \, \Mx &=& V_{x\phantom{'}} \, = \, 
   V_{sys,x\phantom{y'}} \phantom{\cos i_s} + 
   \phantom{(} V_{\sigma,x}   \phantom{ + V_{p,z})  \sin i \, \, \,} + 
   \nonumber\\
   && \phantom{V_{y'} = \, }
   \phantom{(}
   V_{c,x} + V_{p,x} \phantom{)}
   \label{eqn:Vx_parts} \\
\kappa D \, \Myp &=& V_{y'} \, =\,  
    V_{sys,ry'}        \cos i_s + (V_{\sigma,z} + V_{p,z})  \sin i +
    \nonumber\\
    && \phantom{V_{y'} = \, }
   (V_{c,y} + V_{p,y} + V_{\sigma,y}) \cos i
   \label{eqn:Vyp_parts} \\
 && V_{r\phantom{'}} \, = \,
    V_{sys,ry'}        \sin i_s - 
   (V_{\sigma,z} + V_{p,z})  \cos i   +
    \nonumber\\
    && \phantom{V_{y'} = \, }
   (V_{c,y} + V_{p,y} + V_{\sigma,y}) \sin i \, ,
   \label{eqn:Vr_parts}
\end{eqnarray}
where $V_{sys,ry'}$ is the component of the systemic motion in the
plane spanned by the radial and minor axis directions. Note that the
inclusion of the $V_{sys}$ terms ensures that the velocity-gradient
effects [\citet{AG2000}; and \S~\ref{sec:Velocity_Field_Methods}] are
taken into account self-consistently.  These relations involve three
observables ($\Mx$, $\Myp$ and $V_r$) and 3x3+5=14 unknowns (3
$V_{sys}$ terms, 3 $V_\sigma$ projections, 3 $V_p$ terms, $V_c$,
radius $R$, the azimuthal angle $\theta$, distance and
inclination). Furthermore, all observables are expressed with respect
to a coordinate system specified by four additional unknowns: galaxy
centre ($x_0,y_0$), position angle of the major axis ($\phi$) and
height ($z$) above the midplane of the target galaxy\footnote{
Although $z$ does not appear explicitly in
equations~(\ref{eqn:Vx_parts}) -- (\ref{eqn:Vr_parts}), it is present
through the relation between the observable position $y'$ and $y$: $y'
= y \cos{i} + z \sin{i}$}.
Two more observables are available in the position ($x,y'$) of the
star.  We end up with {\em 5 observables} and {\em 18
unknowns}\footnote{
We could add three more unknowns to handle any radial gradients in
inclination, position angle and rotation speed.}.

Obviously, this set of equations cannot be solved. However,
substantial simplifications can be made when ``averaging'' over
several stars. First, the velocity dispersion terms need not be
determined for individual star but only for the group as a whole, so
that only 15 unknowns remain ({\em 3 unknowns shared by all
objects}). Second, all stars share the same systemic motion, so that
the $V_{sys}$ terms follow from ``averaging'' suitably located objects
({\em 3 unknowns shared by all objects}). Third, stars at similar
distance from the galaxy centre share the same rotation speed,
inclination, origin and orientation of the coordinate system ({\em 5
unknowns shared by all objects}), and fourth, the distance is
approximately constant\footnote{
The nearby RP target galaxies are so close that they extend a
substantial depth along the line of sight: $\sim$7\%, $\sim$3.3\%, and
$\sim$2\%, for the LMC, M\,31, and M\,33, respectively. This problem
is solved by replacing $D$ by the sum of the distance to the system
centre ($D_{sys}$) and a position dependent distance:
$D=D_{sys}+d(x,y,z)$. Note that this procedure does not add new
unknowns, just more complexity.}
({\em 1 unknown shared by all
objects}). Equations~(\ref{eqn:Vx_parts}) -- (\ref{eqn:Vr_parts}) can
still not be solved because they involve {\em five observables}, {\em
six star-dependent unknowns} ($\overline{V}_p, R, \theta,$ and $z$) and
{\em twelve shared unknowns} ($x_0, y_0, \phi, \overline{V}_\sigma,
\overline{V}_{sys}, V_c, D,$ and $i$).

To make progress, additional assumptions are needed to reduce the
number of star-based unknowns.  Several approaches are possible. 
One can assume that $z$ and $\overline{V}_p$ lie in the plane of the
circular component as was done by OP2000. This assumption is justified
if the perturbations arise from within the galaxy itself (\eg from the
bar or spiral arms). Thus, $V_{p,z} = 0 =  z$, so that
observations of $N_A$ stars yield $4 N_A$+12 unknown values and $5
N_A$ observables. The assumptions can be generalized by employing simple
parameterizations for $\overline{V}_p$ and the average mid-plane
location $<z>$ as a function of azimuth. For example, the galaxy may
exhibit some corrugation of the plane [as in the Milky\,Way; \eg
\citet{LBH2006}, and references therein] in which case $z$ may be
expressed as a low-order Fourier series with $N_{z,F}$ coefficients,
which behave as shared variables. In the general case with $N_{SV}$
shared variables and four star-based variables, $N_*$ stars yield $4
N_* + N_{SV}$ unknowns and $5 N_*$ observables, leading to the
requirement that:
\begin{eqnarray}
   N_* &\ge& N_{SV}  \, .
   \label{eqn:Nstr_Nsv}
\end{eqnarray}
{\em Thus, observations of $N_{SV}$ stars would suffice to determine
four star-based parameters for each individual star, as well as the
$N_{SV}$ shared unknowns.}  Observing more objects will decrease the
errors of the to-be determined parameters. Note that the equations of
condition are mildly non-linear due to the trigonometric
terms. However, good initial estimates are available for all shared
unknowns, so that an unbiased solution should be attainable via
iteration. This solution scheme is more robust than the simple method
employed by \citet{OP2000}, and will be discussed in more detail
elsewhere \citep{OP2007}.

Employing the solution method outlined above, four of the six
phase-space coordinates can be determined for each observed star,
while two are modelled. Such knowledge for a large number of stars is
likely to be sufficient for the determination of non-circular motions,
as well as the galaxy distance. As noted in the sections on
extra-galactic water masers (\S\S~\ref{sec:Maser_Methods} and
\S\S~\ref{sec:Nuclear_Water_Masers:_Eccentric_Orbits}), the water-maser
method is limited by the fact that it samples only two lines of sight
through the galaxy.  On the other hand, the rotational parallax method
is designed to map the galactic velocity field over a large range in
azimuth with the explicit goal to be able to map out the non-circular
motions. As a result, the RP method will yield more robust distances
than the water-maser method, or, as a matter of fact, any other
proposed method.
\subsubsection{Rotational Parallax: Observational Requirements}
    \label{sec:Rotational_Parallax:_Observational_Requirements}
The distance errors can be reduced by observing more stars, if
systematic errors in the data allow for such averaging.  However,
because non-circular motions can be correlated on large scales, a
substantial number of stars needs to be used to be able to identify
and correct for those systematic non-circular motions. Furthermore,
these stars must be spread out over an area that exceeds the region
affected by the non-circular motions. OP2000 envisaged using stars
spread around an annulus in the target galaxy and estimated that a
minimum of 200 stars are required to achieve a 1\% distance error for
M\,31. Thus, with \SIM\ proper motions for about 200 stars per system,
and if the non-circular motions can be handled properly, the distances
to M\,31, M\,33 and the LMC can be determined to 0.92, 2.0 and 6.4
percent. The required \SIM\ observing time for a rotational-parallax
program depends on the magnitude of the stellar targets. For \GAIA,
with a fixed integration time per star, the final accuracy is
determined by the total number of stars per target galaxy, and the
achieved astrometric accuracy as a function of magnitude. 

If background galaxies and/or QSOs and stars internal to the Local
Group galaxies can be measured simultaneously in the magnitude range
20 -- 24 or so, a pointed, wide-field instrument such as \OBSS\ or
{\em TPF-C} might be well-suited to obtain the required
astrometry. However, for a 1\% distance estimate for M\,31 or M\,33,
the imager needs to be stable to approximately $\mu_{galaxy}/100
\Delta t \sim 0.76 \Delta t$ \muas, or about 3.8 \muas\ for a five
year mission ($\Delta t=5$). This amounts to about 1/11,000\rtp{th} of
a pixel. The \HST\ experience [1/100\rtp{th} of a pixel] indicates
that such accuracies would be hard to accomplish \citep{AK2003}. Thus,
an interferometer such as \SIM\ is the natural instrument to use when
very high accuracies are required. On the other hand, the large
angular extend of the LG galaxies ensures that there are tens of
thousands of usable background galaxies and many more galaxy-member
stars, so that wide-field imagers might be able to the job.

By comparing on-galaxy star counts with off-galaxy counts on fields
with the same Galactic latitude, we estimate that there are a
sufficient number of potential targets in these three local group
galaxies.
%
%
%
%
Based on the UCAC2 catalog \citep{UCAC2}, we estimate that the LMC
contains at least 23,000 stars brighter than V=16 within two degrees
from the centre. Similarly, the 2MASS catalog \citep{2MASS} yields
2,009 (\pmt\ 265) and 984 (\pmt\ 197) stars with K$_s \le 15$ for
M\,31 and M\,33, respectively. The LMC is a natural target for
\GAIA-based RP studies because, at the limiting magnitude of its
radial velocity instrument ($V \sim 17$), \GAIA's proper motion
accuracy is smaller than the LMC's rotational signature of 190
\muasyr.

\citet{Bea2006_H2O} estimate the number of water masers in M\,31 to
equal about 17 at an \SKA\ sensitivity of 1 mJy. Applying their
formalism, we expect a similar number water masers in M\,33. As we
will show below, such small numbers of H$_2$O masers spread out over
the disk of a galaxy are probably not very useful for the
determination of unbiased distances at the percent level. Given the
likely absence of tidal interactions in M\,33, we can probably neglect
$V_{p,z}$ and $z$. Because these masers will be spread out over the
disk, the gradients parameters (for rotation speed, inclination and
position angle) may also be important, so that the M\,33 case may
require 15 shared variables, so that a minimum of 15 water masers are
required [cf. equation~(\ref{eqn:Nstr_Nsv})]. In such situations one
might resort to using the results from the analysis of the neutral
hydrogen (\HI) emission \citep{CS1997}. However, when employing those
rotation-, inclination- and position-angle curves, the solution is no
longer self-consistent.  Furthermore, one may need to apply the
velocity-gradient corrections [\citet{AG2000}; and
\S~\ref{sec:Velocity_Field_Methods}] to interpret the \HI\ data. In
contrast, the large number of bright stars (\apx1000) makes M\,33 a
natural target for \SIM-based RP studies.

It has long been suspected that the interactions of M\,32 and NGC\,205
cause spiral structure and warping in M\,31 [\eg
\citet{gB78,gB83,SK1981,SS1986}]. Recent Spitzer data indicates that
the so-called ``ring of fire'' (where most star formation occurs) is
not centred on M\,31, and that the observed ``split'' in this ring is
likely caused by a recent passage of M\,32 through M\,31's disk
\citep{Gea2006}. Given that most of M\,31's
brightest stars (and hence water masers) are likely to reside in the
ring of fire, and that the large-scale dynamics of these stars/masers
may be significantly affected by tidal interaction, M\,31 may be a
less suitable target for an RP program. Furthermore, the (expected)
small number of water masers is probably not adequate for a reliable
distance measure at the percent level. On the other hand, its large
angular size makes M\,31 a natural candidate for stable, space-based
wide-field imagers such as \OBSS.

A successful dynamical model of the LMC would require at least all the
ingredients uncovered by \citet{vdMea2002}, which include: 1) a
time-dependent inclination, 2) an elliptical disk, 3) low rotation
speed, 4) large internal velocity dispersion, 5) and a significant
bar. In addition, the time-dependency of the position angle of the
line of nodes needs to be considered.  This model is thought to be
sufficient to characterize the tidal effects of the Milky~Way on the
LMC. As such, the LMC may be a less favorable target for a rotational
parallax program. Neglecting these perturbations, the minimum number
of required stars (for a 1\% distance) to overcome just the
velocity-dispersion effects exceeds: $N_{*,LMC} \ge ( 90\%/1\% )^2
= 8,100$. However, given the much larger number of available bright
stars, an accurate rotational parallax for the LMC might still be
achievable with \GAIA.
  \section{Discussion and Conclusions}
\label{sec:Discussion_and_Conclusions}
In this paper we discussed a new, extremely accurate rung of the
extra-galactic distance ladder, namely the rotational-parallax
method. Our preliminary analysis indicates that the best results might
be obtained for the following galaxy/instrument(s) combinations:
M\,33/\SIM, M\,31/\SIM/\OBSS\ and LMC/GAIA/\OBSS. In particular, we
expect that the M\,33-\SIM\ combination can obtain a bias-free
distance with errors no larger than a percent or so. For M\,31/\SIM\
the attainable accuracies are likely to be dominated by the degree to
which the tidal distortions can be modeled. The results for
galaxy/\OBSS-combinations are limited by the level of systematics of
the focal plane array.  For the LMC/\GAIA\ the main uncertainty is the
degree by which localized errors can be beat-down by $\sqrt{N}$
effects.

Assuming $\sqrt{N}$ statistics for the Rotational Parallax method, 2\%
distance errors may be achieved with \SIM-like observations of just 43
stars in M\,31 and 200 stars in M\,33. The LMC requires at least 2,025
stars to reach the desired distance accuracy, mainly because random
motions are significant, so that it is a preferred target for survey
missions such as \GAIA.

An accurate determination of the ``first'' rung of the extra-galactic
distance ladder does not yield the Hubble constant. Distance
indicators that extend well beyond the Local Group must be used for
that. Thus, the RP method would enable a bias-free calibration of the
secondary calibrators. On the other hand, the Water Maser Cosmology
project \citep{WMCP2006} might be better than the RP method in the
short term because it can be already applied to a number of galaxies,
albeit with less accuracy per system.

We expect that \GAIA's delivery of five-dimensional phase-space
coordinates for a magnitude limited sample of individual stars of
external galaxies (LMC, SMC et cetera) will have many significant
galactic-dynamics applications such as bar-dynamics, disc dynamics and
tidal interactions. On the other hand, M\,31 and M\,33 naturally lead
to \SIM-based programs because the internal motions (relative to the
stellar velocity dispersions) are four to ten times smaller than for
the LMC.

In this paper we focussed our cosmological analyses on the case that
\PLANCK\ data will be the norm for CMB data, and other data
(Baryon Acoustic Oscillation, Galaxy Clusters, Supernovae and Weak
Lensing) have accuracies at a level coined Stage~I by the Dark Energy
Task Force \citep{DETF2006}. In such a situation, the effects of
decreasing the error on the Hubble constant on our knowledge of the
equation of state of Dark Energy (and $\Omega_{tot}$) declines when
$H_0$ is known more accurately. For example, with \PLANCK-like CMB
data, decreasing the error on $H_0$ by a factor of two from the
current uncertainty will almost half the error on $w$. However, when
decreasing $\epsilon_{H_0}$ from 2\% to 1\%, $\epsilon_w$ (and
$\epsilon_{\Omega_{tot}}$) decreases by only 20\% (see
Fig.~\ref{fig:Cosmology}).

Scaling our analysis to the situation that the other data set are
much more accurate than at present, we find that the effects of
smaller $\epsilon_{H_0}$ values become gradually less important. In
accordance with the DETF we find that when the quality and quantity of
the other data sets reaches the Stage~IV level, decreasing
$\epsilon_{H_0}$ has only a minor effect.

An alternate utility of determining $H_0$ with great accuracy from
nearby galaxies is that it is fully independent of traditional
cosmological analyses which {\em determine} $H_0$, while employing
only a weak prior of its value. Especially when both methods yield
$H_0$ values with accuracies at the percent level or so, it becomes
possible to test fundamental assumptions underlying CBM-type analyses
such as the validity of general relativity as a description for the
expansion of the Universe \citep{DETF2006}.

Other fundamental test of cosmology will become available via \SIM,
\GAIA\ and \OBSS\ astrometry in the form of determinations of very
accurate ages of stars in (eclipsing) binary systems
\citep{LB2000}. For such systems, all relevant fundamental stellar
parameters (mass, radius, luminosity and metallicity) can be
determined with great accuracy, leaving no wiggle room for stellar
models. The accuracy of stellar dating is fundamentally limited by
luminosity (distance) uncertainty because stellar evolution predicts
luminosity ($L$) evolution of about $\Delta L/L \approx (0.24 -0.106
\log{\tau_{MS}}) \pm 0.05$ per Gyr, where $\tau_{MS}$ is the
main-sequence (MS) lifetime of the star in Gyr [e.g.,
\citet{OFW2003}]. A G6V star (0.89 M$_\odot$) has a MS lifetime roughly
equal to the age of the Universe, while its luminosity evolution
amounts to about +12\% per Gyr. If one-percent distances (2\% in $L$)
are available for such stars in the Galactic halo, then a measurement
of their apparent magnitude corresponds to a determination of their
age to approximately $1000\times (2\%/12\%)$ = 166 Myr.  Such binary
star observations will also yield the stars' interior Helium abundance
($Y$), so that $Y(t)$ can be determined and extrapolated to
$t=0$. Many thousands of such halo stars are accessible by \GAIA,
\SIM\ and \OBSS, so that a very accurate lower limit of the age of the
Universe can be obtained right here in the Solar neighbourhood. If
one-percent distances are available for nearby Local Group systems,
such accurate age determinations can potentially be made in those
galaxies, so that a wide range in star-formation histories can be
sampled.

\vspace*{2em}
\nid I thank Dean Peterson, David Spergel, Ed Shaya and Alan Peel for
useful discussions. I am also grateful for several suggestions by the
referee to improve the paper, especially with regard to Hu's paper and
the need to distinguish between an accurate extra-galactic distance
and an accurate Hubble constant.

\clearpage

\onecolumn

\clearpage
\twocolumn

\appendix
\section[]{Water Masers \& Elliptical Orbits}
\label{apx:elliptical_orbits}
For the general case of elliptical orbits, Newtonian gravity and a
dominant central object, the orbital speed at distance $R$ is given
by:
\begin{eqnarray}
   V_o(R) &=& \sqrt{ \frac{G \, M_o}{a_o} \, \frac{2 - \alpha}{\alpha} }
   \label{eqn:V_orbit}\\
   V_{o;km/s}(R) &=&   2 \pi \kappa
   \sqrt{ \frac{M_{o}}{a_{o;AU}} \, \frac{2 - \alpha}{\alpha} } 
   \label{eqn:V_orbit_kms}
\end{eqnarray}
where $G$ is the gravitational constant, $M_o$ the mass of the central
object, $a_o$ the semi-major axis of the orbit, and $\alpha = R/a_o$.
The subscripts ``$c$,'' ``$e$,'' and ``$o$'' indicate cases that are
valid for circular, elliptical and general orbits, respectively. We
also specify the units of the parameter in the subscript. For
eqn.~(\ref{eqn:V_orbit_kms}) we express the velocity in \kms, with the
mass of the central object always in units of $M_{\sun}$, $\kappa$
(\apx 4.74) converts velocities in AU yr\rtp{-1} to \kms\ and
$a_{o;AU}$ equals $a_o$ in astronomical units.  For circular orbits
($\alpha=1$), eqn.~(\ref{eqn:V_orbit}) reduces to the well-known
$V_o(R) = \sqrt{G \, M/a_o}$, while for orbits with eccentricity $e$
at we have $V_{o;peri} = \sqrt{G \, M/a_o \times (1+e)/(1-e)}$ at
peri-centre and $V_{o;apo} = \sqrt{G \, M/a_o \times (1-e)/(1+e)}$ at
apo-centre.

For our test geometry (see
\S\S~\ref{sec:Nuclear_Water_Masers:_Eccentric_Orbits}), the
high-velocity masers are located at the tangent of the ellipse at
distance $R_{HV} = \sqrt{b^2 + (e\,a_e)^2} =
\sqrt{ (1-e^2)\, a_e^2 + (e\,a_e)^2} = a_e$ from the centre, where $b$
is the semi-minor axis of the ellipse. Thus, for the high-velocity
masers, $\alpha$ equals unity, so that their observed radial
velocities map out a rotation curve that is identical to the case of
circular orbits.

To derive the expected kinematics, we use a coordinate system with $x$
and $y'$ axes that are aligned with the apparent major and minor axes
(neglecting the position-angle warp), and with $y$ the corresponding
coordinate in the plane of the circum-nuclear disk (i.e., $y' = y
\cos{i}$).  Because the maser spots move only a small amount in azimuth
during a multi-year experiment, we can approximate the orbital motion
as being locally sinusoidal with constant angular velocity, and with
period $P \approx 2\, \pi R / V$. Then, for Heliocentric
distance $D$, the positions, radial velocities ($V_r$), radial
acceleration ($\dot{V}_r$) and proper motion ($\mu$) are:
\begin{eqnarray}
x       &\approx&   R_{o} \, \sin{( \frac{2\pi t}{P_{o}} )} 
      \, = \, R_{o} \, \sin{( \frac{t \, V_{o}}{R_{o}} )}
   \label{eqn:x_H2O}\\
y       &\approx& R_{o} \, \cos{( \frac{t \, V_{o}}{R_{o}} )}
   \label{eqn:y_H2O}\\
\frac{V_{r,HV;o}}{\sin{i_{HV;o}}} \hspace*{-0.5em} &=& \hspace*{-0.5em}  
    \dot{y}_{HV} \, = \, V_{HV;o}
   \label{eqn:V_HV}\\
\frac{\abs{V_{r,S;o}}}{\sin{i_{S;o}}} \hspace*{-0.5em} &=& \hspace*{-0.5em}  
   \dot{y}_{S;o}
      \, = \, V_{S;o}\phantom{^2} 
           \sin{(\frac{t \, V_{S;o}}{R_{S;o}})}
        \, \approx \, \frac{V_{S;o}^2}{R_{S;o}} \, t
   \label{eqn:V_re} \\
\frac{\abs{\dot{V}_{r,S;o}}}{\sin{i_{S;o}}} \hspace*{-0.5em} &=& \hspace*{-0.5em} 
   \ddot{y}_{S;o} \, = \, 
   \frac{V_{S;o}^2}{R_{S;o}} \cos{(\frac{t \, V_{S;o}}{R_{S;o}}) } 
        \, \approx \, \frac{V_{S;o}^2}{R_{S;o}}
   \label{eqn:A_re}\\
\mu_{S;o} \hspace*{-0.5em} &=& \hspace*{-0.5em} 
   \frac{ \dot{y}_{S;o} } { \kappa D } 
   \phantom{\, = \,
   \frac{V_{S;o}^2}{R_{S;o}} \cos{(\frac{t \, V_{S;o}}{R_{S;o}}) }
   }
   \, \approx \, \frac{V_{S;o}}{\kappa D_{S;o}} \, ,
   \label{eqn:V_te}
\end{eqnarray}
where we use the additional subscripts ``S'' and ``HV'' to indicate
whether the expression is valid for the systemic or high-velocity
masers, respectively. The units of the proper motion are \masyr\ if
the velocities are in \kms\ and the distance in kpc.

First, we will examine the constraints that derive from the
high-velocity maser spots. For our toy model, the velocities in
eqns.~(\ref{eqn:V_orbit_kms}) and (\ref{eqn:V_HV}) are equal so that
the observable radial velocity is:
\begin{eqnarray}
  V_{r,HV;o} &=&  2 \pi \kappa \, \sin{i_{HV;o}} 
     \sqrt{   \frac{ M_{o} \, 1000D_{o;Mpc} }{ a_{HV;o;mas} } } \, ,
  \label{eqn:HV_Observable}
\end{eqnarray}
where $a_{HV;o;mas}$ is the observed major-axis position of the
high-velocity spot. Bringing all the observables to the left-hand side
(LHS) yields: 
\begin{eqnarray}
  \frac{ a_{HV;o;mas} V^2_{r,HV;o}}{4000 \pi^2 \kappa^2} &=&  
  \sin^2{i_{HV;o}} M_{o} D_{o;Mpc} \, .
  \label{eqn:HV_Observable_LHS_RHS}
\end{eqnarray}
Turning our attention to the systemic masers, we use the observed
minor-axis position of the systemic maser spots to determine an
expression for the semi-major axis. First, we have: $y'_{S;o;AU} =
R_{S;o;AU} \cos{ i_{S;o}} = a_{o;AU} (1-e_o) \cos{i_{S;o}} =
y'_{S;o;mas} \times (1000 \, D_{o;Mpc})$, so that:
\begin{eqnarray}
a_{S;o;AU} &=& \frac{ y'_{S;o;mas} \times (1000 \, D_{o;Mpc})}
                  { (1-e_o) \cos{ i_{S;o} }                 }
   \label{eqn:a_o_AU_mas} \\
   \frac{ a_{S;o;mas} }{y'_{S;o;mas} } &=& \frac{1}{(1-e_o) \cos{ i_{S;o} } }
   \, .
   \label{eqn:yobs_sys}
\end{eqnarray}
where the second relation has all observables on the LHS. Inserting
eqn.~(\ref{eqn:a_o_AU_mas}) into (\ref{eqn:V_orbit_kms}) and the
result into (\ref{eqn:V_te}), we find:
\begin{eqnarray}
   \frac{ 10^9 }{ 4 \pi^2 } \, y'_{S;o;mas} \times \mu^2_{S;o;mas/yr}
     \hspace*{-0.5em} &=& \hspace*{-0.5em}
     \frac{ M_{o} \cos{i_{S;o}} (1+e_o) }
          { D^3_{o;Mpc}                      } \, .
   \label{eqn:yp_x_mu^2}
\end{eqnarray}
Further information is contained in the observed acceleration of the
systemic maser spots: inserting eqn.~(\ref{eqn:V_orbit_kms}) evaluated
at peri-centre into eqn.~(\ref{eqn:A_re}) and using
(\ref{eqn:a_o_AU_mas}) we obtain:
\begin{eqnarray}
   \frac{\dot{V}_{r;S;o} (1000 y'_{S;o;mas})^2}{4\pi\kappa^2} 
   \hspace*{-0.7em} &=& \hspace*{-0.7em} 
      \frac{ \sin{i_{S;o}} \cos^2{i_{S;o}} \, M_{o} \, (1+e_o)}
           { D^2_{o;Mpc}                                         } \, ,
   \label{eqn:Vr_dot}
\end{eqnarray}
If one assumes circular orbits, a constant inclination and that
$a_{S}=a_{HV}$, then there are two sets of three equations
[[(\ref{eqn:HV_Observable_LHS_RHS}), (\ref{eqn:yobs_sys}) and
(\ref{eqn:yp_x_mu^2})] and [(\ref{eqn:HV_Observable_LHS_RHS}),
(\ref{eqn:yobs_sys}) and (\ref{eqn:Vr_dot})]] that can be used to
solve for the three unknowns: distance, inclination and the mass of
the central object.  If a simple warped model [$i = i_0 + \beta
(R-R_0)$] is adopted, one can determine the slope ($\beta$) from the
high-velocity maser spots because they have a warped (projected)
geometry and a small non-Keplerian behavior \citep{HGM1996}.

Equations (\ref{eqn:HV_Observable_LHS_RHS}), (\ref{eqn:yobs_sys}),
(\ref{eqn:yp_x_mu^2}) and (\ref{eqn:Vr_dot}) can be rearranged into
the following relations between the circular and eccentric cases:
\begin{eqnarray}
   \sin{i_{HV;c}} \sqrt{ M_{c} D_{c;Mpc} }
      \hspace*{-0.5em} &=& \hspace*{-0.5em}
   \sin{i_{HV;e}} \sqrt{ M_{e} D_{e;Mpc} }
   \label{eqn:HV_constraints}\\
   \cos{ i_{S;c} } &=& (1-e) \cos{ i_{S;e} }
   \label{eqn:SysPos_constraints}\\
   \frac{ M_{c} \cos{i_{S;c}}      } { D^3_{c;Mpc} } &=& 
   \frac{ M_{e} \cos{i_{S;e}} (1+e)} { D^3_{e;Mpc} }
   \label{eqn:MuS_constraint}\\
   \frac{ \sin{i_{S;c}} \cos^2{i_{S;c}} \, M_{c}}
        { D^2_{c;Mpc}                           }
   \hspace*{-0.8em} &=& \hspace*{-0.8em} 
   \frac{ \sin{i_{S;e}} \cos^2{i_{S;e}} \, M_{e} \, (1+e)}
        { D^2_{e;Mpc}                                    } \, ,
   \label{eqn:dVrS_constraint}
\end{eqnarray}
where constraint equations (\ref{eqn:HV_constraints}) --
(\ref{eqn:dVrS_constraint}) also equal the LH sides of equations
(\ref{eqn:HV_Observable_LHS_RHS}), (\ref{eqn:yobs_sys}),
(\ref{eqn:yp_x_mu^2}) and (\ref{eqn:Vr_dot}), respectively.  Again we
can include a fifth relation from the warping behavior, but for
illustrative purposes we will assume a constant inclination, so that
we have four equations [(\ref{eqn:HV_constraints}) --
(\ref{eqn:dVrS_constraint})] with four unknowns [$D, i, M$ and
$e$]. Expressing the first three unknowns as a function of the fourth,
we find:
\begin{eqnarray}
\frac{\cos{ i_{e}}}{\cos{ i_{c}}} &=&
    \phantom{\sqrt{(2)} }
     \frac{1}{1-e} \, \approx \, 
          1 + \phantom{2\,}e + 2\,e^2 + {\cal O}(e^3)
   \label{eqn:i_ell}\\
\frac{ D_{e}}{D_c} &=& \sqrt{ \frac{1+e}{(1-e)^3} } \, \approx \, 
          1 + 2\,e + \frac{5}{2} \, e^2 + {\cal O}(e^3)
   \label{eqn:D_ell}\\
\frac{M_{e}}{M_c} &=& \sqrt{ \frac{1+e}{(1-e)^7} } \, \approx \, 
         1 + 4\,e + \frac{19}{2} \, e^2 + {\cal O}(e^3) \, ,
   \label{eqn:M_ell}
\end{eqnarray}
where we note that these relations hold for systemic masers at
peri-centre. One should keep in mind that eqns.~(\ref{eqn:i_ell}) --
(\ref{eqn:M_ell}) were derived for the special geometry described in
\S\S~\ref{sec:Nuclear_Water_Masers:_Eccentric_Orbits}.
These simplifications were used to illustrate the effects of
unmodeled eccentric orbits for nuclear water masers, not to provide
accurate relations for any specific case.

Above we have worked out the case that the systemic masing occurs at
peri-centre. For the alternate case (masing at apo-centre), we find
that we can replace $e$ with $-e$ in equations (\ref{eqn:a_o_AU_mas})
-- (\ref{eqn:M_ell}), including the approximations of
(\ref{eqn:i_ell}) -- (\ref{eqn:M_ell}).

\end{document}